\documentclass[journal=jpccck,manuscript=article]{achemso}
\usepackage[version=3]{mhchem} 
\usepackage{graphicx}
\usepackage{epsfig}
\usepackage{multirow}
\usepackage{dcolumn}

\usepackage{bm}
\usepackage{subfigure}
\usepackage{color}

\author{Vijay Kumar Gudelli}
\author{V. Kanchana}
\email{kanchana@iith.ac.in}
\affiliation{Department of Physics, Indian Institute of Technology Hyderabad, Ordnance Factory Estate, Yeddumailaram-502 205, Andhra Pradesh, India}
\author{S. Appalakondaiah}
\author{G. Vaitheeswaran}
\affiliation{Advanced Center of Research in High Energy Materials (ACRHEM), University of Hyderabad, Prof. C. R. Rao Road, Gachibowli, Hyderabad-500 046, Andhra Pradesh, India}
\author{M. C. Valsakumar}
\affiliation{School of Engineering Sciences and Technology (SEST), University of Hyderabad, Prof. C. R. Rao Road, Gachibowli, Hyderabad-500 046, Andhra Pradesh, India}
\date{\today}

\title[An \textsf{achemso} demo]
  {Phase Stability and Thermoelectric Properties of the Mineral FeS$_2$: An Ab-initio Study}

\begin{document}

\begin{abstract}
  First principles calculations were carried out to study the phase stability and thermoelectric properties of the naturally occurring marcasite phase of FeS$_2$ at ambient condition as well as under pressure. Two distinct density functional approaches has been used to investigate the above mentioned properties. The plane wave pseudopotential approach was used to study the phase stability and structural, elastic, and vibrational properties. The full potential linear augment plane wave method has been used to study the electronic structure and thermoelectric properties. From the total energy calculations, it is clearly seen that marcasite FeS$_2$ is stable at ambient conditions, and it undergoes a first order phase transition to pyrite FeS$_2$ at around 3.7 GPa with a volume collapse of about 3$\%$. The calculated ground state properties such as lattice parameters, bond lengths and bulk modulus of marcasite FeS$_2$ agree quite well with the experiment. Apart from the above studies,  phonon dispersion curves unambiguously indicate that marcasite phase is stable under ambient conditions. Further, we do not observe any phonon softening across the marcasite to pyrite transition and the possible reason driving the transition is also analyzed in the present study, which has not been attempted earlier. In addition, we have also calculated the electronic structure and thermoelectric properties of the both marcasite and pyrite FeS$_2$. We find a high thermopower for both the  phases, especially with p-type doping, which enables us to predict that FeS$_2$ might find promising applications as good thermoelectric materials.
\end{abstract}
\clearpage

\section{Introduction}
Experimental and theoretical studies of iron based sulphides have been of considerable interest in the last few decades, in the geophysical context, in view of their occurrence in marine systems and surface of earth as well as other planetary systems\cite{Errandonea}. These compounds exhibit a wide range of interesting structural characteristics such as epitaxial inter-and over-growth of their polymorphic phases at lower temperatures. Also, they are found to exist in a variety of  iron or sulphur deficient forms, which has implications on their mining and geochemical processing \cite{Rieder,Kjekshus,Donohue,Takizawa,Yamashita,Goodenough,Hull,Gou}.
Among the iron based sulphides, FeS$_2$ is the most abundant natural mineral, and it is available in two closely related polymorphic structures, {\it viz.},marcasite and pyrite. Marcasite FeS$_2$ is the commonly available mineral in hydrothermal systems and in  sedimentary rocks, whereas pyrite FeS$_2$ is the most abundant mineral on the earth's surface. Several works have been reported in the past to investigate the similarities and differences between marcasite and pyrite crystal structures \cite{David,Istvan,Peter}. In both the structures, the Fe atoms are octahedrally coordinated with six S atoms and the S atoms are in tetrahedral coordination with three Fe atoms and one S atom. In detail, marcasite FeS$_2$ crystallizes in the orthorhombic structure (space group $Pnnm$, Z=2) with Wyckoff position $2a (0, 0, 0)$ for Fe, and $4g (u, v, 0)$ for S. The Sulphur octahedra, with which the Fe atoms are octahedrally coordinated, share the $\left[1 1 0\right]$ edges- see the crystal structure as shown in Fig. 1(a).  On the other hand, pyrite FeS$_2$ crystallizes in cubic structure (space group $Pa\bar{3}$, Z=4) with Wyckoff position $4a (0, 0, 0)$ for Fe and  $8c (u, u, u)$ for S. Thus in the pyrite structure, the Fe atoms are situated at the corners and face centers, whereas the S atoms are placed in the form of S-S dumbbells oriented along $\left<111\right>$ directions with their centers at the body center and edge centers of the cube. A number of experiments have been performed on bulk pyrite and also on surfaces aiming to study the spectroscopic properties due to their potential applications in photo-voltaic industry
\cite{Murphy,Ennaoui,Ferrer,Wadia,Wu1,Spagnoli,Birkholz,Bronold,Oertzen,Halim,Oertel,Mamiya}.  Jagadessh and Seehra performed electrical resistivity measurements on natural marcasite and its energy gap was found to be 0.34 eV \cite{Jagadesh}. The dynamical properties of these compounds have been measured by different experimental and theoretical methods \cite{Spagnoli,Lutz,sourisseau,kleppe,buhreer}. Apart from these experiments, several theoretical studies have also been carried out to explore the structural properties of FeS$_2$ using  combinations of Density Functional Theory (DFT), Gaussian/Plane-wave  basis sets and all-electron/pseudopotential approaches. The mechanical properties of pyrite FeS$_2$ have been explored at different pressures using classical inter-atomic potentials and marcasite is predicted to react in a manner similar to pyrite when compressed \cite{Sithole1}. The electronic structure calculations for FeS$_2$ were performed using non empirical atomic orbital method and the band gaps are reported for pyrite and marcasite forms of FeS$_2$ \cite{Bullett}. Ahuja {\it et al} \cite{Ahuja} has given a detailed discussion about the interband transitions in pyrite FeS$_2$ through optical properties. Merkal {\it et al} \cite{Merkel} performed both experimental and theoretical investigations and concluded that cubic symmetry exists under high pressure conditions.

\par Albeit availability of numerous studies that analyze pyrite and marcasite from various perspectives, a precise study on both the polymorphic forms of FeS$_2$ is needed to explain the phase stability of FeS$_2$ at ambient conditions. In the present work, we have studied both the polymorphic forms of FeS$_2$ and have accounted for the first order phase transition between marcasite and pyrite at high pressure and at zero temperature accompanied by a reasonable volume collapse which was not shown in any of the earlier studies and we have also speculated on the reasons driving the transition. Ruoshi Sun {\it et  al} \cite{Ruoshi} have also studied the relative stability of FeS$_2$ with the aim to explore the photo-voltaic performance of FeS$_2$. Similarly, it is also interesting to note that FeS$_2$ possesses a fairly good thermopower\cite{Bither}, which has not been addressed theoretically till now. Thermoelectric materials (TE) find potential applications including power generation, refrigeration, and had been a thrust area of research for the experimentalists and theoreticians for the past few decades, provoking their thoughts in search of a material with better performance. TE materials can convert waste heat into electric power and hence they can play a vital role in meeting the present energy crisis and environment pollution \cite{David1,YuLi}. The performance of a TE material is quantified by the dimensionless figure of merit ZT which is given by ZT=S$^2$$\sigma$T/$\kappa$, where S, $\sigma$, T, $\kappa$ are the Seebeck coefficient, electrical conductivity, absolute temperature and the thermal conductivity (which includes both the electronic $\kappa$$_e$ and lattice contribution $\kappa$$_l$. {\it i.e.} $\kappa$=$\kappa$$_e$+$\kappa$$_l$) respectively.  It is clear that the value of ZT can be increased by making the values of thermo power and electrical conductivity high while keeping  the value of thermal conductivity low. Assuming that the value of $\kappa$ can be reduced to the amorphous limit, a way of maximizing ZT is to maximize $S^2\sigma$. We have calculated the thermoelectric properties of FeS$_2$ in both the polymorphic phases for the first time in this perspective.
\par The rest of the paper is organized as follows: In section II we describe the method used for the theoretical calculations and structural properties are presented in section III. In section IV, we discuss the electronic structure and thermoelectric properties are presented in section V and conclusions are given in section VI.

\section{Method of calculations}
	All the total energy calculations were performed using Plane wave self-consistent field (Pwscf) program based on  density functional theory, plane wave basis set and pseudo potential method \cite{Pwscf}.  The total energies are obtained by solving the Kohn-Sham equation self consistently within the Generalized Gradient Approximation (GGA) of Perdew-Burke-Ernzerhof (PBE) potential \cite{Perdew}. A plane wave kinetic energy cut off of 50 Ry is used and the first Brillouin zone  is sampled according to the Monkhorst-Pack scheme \cite{Monkhorst} by means of a $8\times 8\times 8$ k-mesh in order to ensure well converged results.  The electron-ion interactions are described by Vanderbilt type ultrasoft pseudo potentials \cite{Vanderbilt} and the pseudo potentials are treated with nonlinear core corrections with the following basis set Fe: 3s$^2$ 3p$^6$ 3d$^6$ 4s$^2$ and S: 3s$^2$ 3p$^4$ as valence states. The  variable-cell structural optimization has been performed using Broyden-Fletcher-Goldfarb-Shanno (BFGS) algorithm as implemented in Pwscf.  In order to obtain information about the relative phase stability of marcasite and pyrite phases as a function of pressure, we have calculated the cohesive energy of both the phases at pressures ranging from -3 GPa (expansion) to 9 GPa (compression) with a step size of 0.5 GPa. For each pressure, structural optimization of the unit cell has been carried out by relaxing positions of all the atoms together with the necessary changes in the shape and volume. The threshold criteria of $1\times 10^{-5}$ Ry for total energies, $1\times 10^{-4}$ Ry/bohr for the maximum force and 0.002 GPa for total stress were used for the structural relaxation. The phonon dispersion calculations are performed within the frame work of Density Functional Perturbation Theory (DFPT). Dynamical matrices were setup and diagonalized for phonon wavevectors corresponding to a Monkhorst-Pack grid of $4\times 4\times 4$. In order to calculate the elastic properties of FeS$_2$, we have used CASTEP package \cite{Payne,Segall} which is also based on the plane wave pseudopotential method. To obtain well converged parameters, we have used a plane wave cut off of 800 eV and Monkhorst-Pack \cite{Monkhorst} grid  with a minimum spacing of 0.025 \AA$^{-1}$ using GGA-PBE exchange correlation functional with ultrasoft pseudopotential. The elastic constants can be computed by calculating the elastic energy as a function of elastic strain. More specifically, the curvature of the elastic energy as a function of value of a particular type of elastic strain gives the value of a particular combination of elastic constants. By repeating the calculations for adequate number of independent strains, we can obtain the required number of independent equations for the elastic constants which can be solved to obtain the values of the independent elastic constants.  For each strain, the coordinates of the ions are fully relaxed keeping the shape of the unit cell corresponding to the given strain intact. After obtaining the single crystal elastic constants, we have calculated polycrystalline properties such as the bulk modulus, shear modulus, sound velocities and Debye temperature using the Voigt-Ruess-Hill (VRH) \cite{Hill} approximation- see Ravindran {\it et al} \cite{Ravindran} for a  detailed discussion on these calculations.

\par To study the electronic properties, we have used full-potential linear augmented plane wave (FP-LAPW) method based on first-principles density functional calculations as implemented in the WIEN2k \cite{Blaha}. As it is well known, for the semiconductors and insulators,  the electronic band gap calculated using DFT with the standard exchange-correlation functionals such as LDA and GGA  is usually   about 30 to 40\% less when compared to experiments due to self-interaction and lack of the derivative discontinuities of the exchange correction potential with respect to occupation number\cite{Panchal,Tran}.
In the present work, we have used a modified GGA, known as Tran and Blaha modified Becke Johnson potential (TB - mBJ)\cite{Becke}, which is found to be quite successful in reproducing the experimental band gaps as compared to standard GGA \cite{Tran,Koller1,Koller2,Dixit,Hong}. Here we have used $9\times 8\times 12$ and  $15\times 15\times 15$ Monkhorst-Pack k-meshes for the self-consistent calculations, resulting in 175 and 176 k-points in the irreducible part of the Brillouin zone, respectively for the marcasite and pyrite phases. Spin-orbit coupling has been incorporated in our calculations. All our calculations are performed using the optimized parameters from the PWscf calculation with an energy convergence up to 10$^{-6}$ Ry per unit cell between the successive iterations. Further, we have calculated the thermopower (S), and $\frac{\sigma}{\tau}$  using BOLTZTRAP\cite{Madsen} code  with well converged (using as many as 100000  k-points in the Brillouin zone) using the self-consistent calculation, with in the Rigid Band Approximation (RBA) \cite{Scheidemantel,Jodin} and the constant scattering time ($\tau$) approximation (CSTA). According to the RBA approximation, doping a system does not vary its band structure but varies only the chemical potential, and it is a good approximation for doped semiconductors to calculate the transport properties theoretically when doping level is not very high \cite{Jodin,Chaput,Bilc,Ziman,Nag}. According to CSTA, the scattering time of the electron is independent of energy and depends on concentration and temperature. The detailed explanation about the CSTA is given in Ref \citenum{singh, aggate2,  Khuong}. and the references therein. It is evident that CSTA had been quite successful in predicting the thermoelectric properties of many materials in the past \cite{DJS,Parker,Lijun}.

\section{Results and discussion}
\subsection{Phase stability of FeS$_2$}

The structural phase stability of the marcasite and pyrite phases of FeS$_2$ has been studied to find the ground state of FeS$_2$ at various pressures. As a first step, we have performed total energy calculations as a function of volume for both marcasite and pyrite FeS$_2$ structures by varying pressures from -3 GPa to 9 GPa. Here, the negative and positive pressures represent respectively expansion and compression of the unit cell. As shown in Fig. 2, the total energy curves clearly show marcasite to be energetically favorable than pyrite with an energy difference of $\sim$0.03 Ry/atom. A similar result has been obtained by a recent study on FeS$_2$ using a different numerical implementation \cite{Spagnoli}. Even though similar studies exist in the literature, none of them clearly identified marcasite as the ground state of the polymorphic FeS$_2$.  From Fig. 2, we found a possible structural phase transition from marcasite to pyrite FeS$_2$ $\sim$0.98 of V/V$_0$ where V is the theoretical volume and V$_0$ is the experimental volume. To obtain the transition pressure, we have also plotted the enthalpy difference versus pressure which is presented in Fig. 3 (a). From this, the pressure for marcasite to pyrite phase transition is determined to be 3.7 GPa. As shown in Fig. 3 (b), there is  a volume collapse $\sim 3\%$ which indicates the first order character of the phase transition from marcasite to pyrite structure.  In the present study we specifically address the ground state as well as high pressure properties of marcasite FeS$_2$ which was not reported previously. It is interesting to note that there are no conspicuous signatures of the marcasite to pyrite transition in the phonon dispersion, Raman or IR spectra of marcasite FeS$_2$ at ambient as well as high pressures. We shall discuss this in more detail in later sections.

\subsection{Structural properties}

The calculated structural properties such as lattice constants, volume, internal parameters (u and v) of sulphur and bond lengths of Fe-Fe, Fe-S and S-S for marcasite FeS$_2$ are compared with experiments and other theoretical reports and are summarized in Table 1. The calculated lattice parameters and volume are in reasonable agreement with the calculations of Spagnoli {\it et al} \cite{Spagnoli} and Sithole {\it et al} \cite{Sithole}. The minor differences found when compared with the results of  Sithole {\it et al} \cite{Sithole} are due to different parameterizations of the exchange-correlation functional used in their calculations.

\begin{table}[ht]
\caption{Ground state properties of marcasite FeS$_2$ at ambient pressure combined with other experimental and theoretical reports.}
\begin{tabular}{lllllllllll}

Parameters    &      &Present work	&     	&Exp.         &      &results from other calculations \\\hline

a (\AA)       &      &4.439          	&	&4.436$^a$    &      &4.434$^c$, 4.373$^d$ \\

b (\AA)       &      &5.408            	&	&5.414$^a$    &      &5.404$^c$, 5.381$^d$  \\

c (\AA)       &      &3.388            	&	&3.381$^a$    &      &3.387$^c$, 3.407$^d$  \\

u (S)         &      &0.206            	&	&0.200$^a$    &      &0.203$^d$         \\

v (S)         &      &0.375            	&	&0.378$^a$    &      &0.380$^d$  \\

V (\AA$^3$)   &      &81.33            	&	&81.20$^a$    &      &81.16$^c$, 80.17$^d$ \\

B$_0$ (GPa)   &      &150.1            	&	&146.5$^b$    &                        \\

B'            &      &5.4              	&	&4.9$^b$      &                          \\

Fe-Fe (\AA)   &      &3.38	       	&	&3.36$^a$     &     &3.386$^d$   \\

Fe-S (\AA)    &      &2.23	       	&	&2.21$^a$     &     &2.229$^d$    \\

S-S (\AA)     &      &2.20             	&	&2.19$^a$     &     &2.195$^d$    \\
\hline
\end{tabular}\\
$a$ : Ref. \citenum{Buerger};
$b$ : Ref. \citenum{Chattopadhyay};
$c$ : Ref. \citenum{Spagnoli};
$d$ : Ref. \citenum{Sithole}
\end{table}

\par The relative change of lattice parameters a, b and c with respect to external pressure upto 5 GPa are shown in Fig. 4 (a). From this we found that, marcasite FeS$_2$  is relatively less compressible along $\vec{b}$-direction than along $\vec{a}$ and $\vec{c}$ directions. The  pressure coefficients y(x)= $\lvert{\frac{1}{x}\frac{dx}{dP}\rvert}_{P=0}$ (with x being either a, b or c) of the lattice parameters are found to be $2.2\times 10^{-3}$, $1.8\times 10^{-3}$ and $1.9\times 10^{-3}$ and  GPa, respectively so that y(b)  $<$ y(c) $<$ y(a) which implies least compressibility along the $\vec{b}$-axis. The Fe-S$_i$ (i =1,2)bond lengths monotonically decrease with pressure as illustrated in Fig. 4(b). The variation of fractional coordinates (u and v) of sulphur as a function of pressure is found to be less significant.  Overall, the structural parameters of marcasite FeS$_2$ show minor changes under pressure, which implies that marcasite FeS$_2$ shows almost isotropic behavior with external pressure despite crystallizing in orthorhombic structure. The bulk modulus(B$_0$) and its pressure derivative(B'), calculated using Birch-Murnaghan equation of state, are 150.1 GPa and 5.7 respectively for marcasite FeS$_2$  which are in good agreement with the earlier experimental values of 146.5 GPa and 4.9, respectively \cite{Chattopadhyay}.

\subsection{Mechanical properties}
\par To obtain mechanical stability of marcasite type FeS$_2$, we have calculated elastic properties of this material. Elastic constants are the fundamental material parameters that describe the resistance of the material against applied mechanical deformation. Since marcasite FeS$_2$ crystallizes in orthorhombic structure, it has nine independent elastic constants namely C$_{11}$, C$_{22}$, C$_{33}$, C$_{44}$, C$_{55}$, C$_{66}$, C$_{12}$, C$_{13}$ and C$_{23}$. To calculate the elastic constants, we have performed complete structural optimization of the  experimental structure using CASTEP. The calculated single crystal elastic constants at the theoretical equilibrium volume are tabulated in Table 2. All the calculated single crystal elastic constants satisfied the Born's mechanical stability criteria for orthorhombic structure \cite{Born}, thereby implying that the marcasite type  FeS$_2$  is mechanically stable under ambient conditions. From the calculated values it is clearly observed that C$_{22}$ $>$ C$_{33}$ $>$ C$_{11}$ which implies that marcasite FeS$_2$ is  stiffer along $\vec{b}$-direction than along $\vec{c}$ and $\vec{a}$ directions. The single crystal bulk modulus calculated from elastic constants is 164.8 GPa, which is in reasonable agreement with the value 150.1 GPa obtained in this work using Birch-Murnaghan equation of state. By using the calculated single crystal elastic constants, we further computed the polycrystalline aggregate properties such as Bulk moduli (B$_X$, X=V, R, H), Shear moduli (G$_X$,  X=V, R, H) using the Voigt, Reuss and Hill approaches. The calculated  polycrystalline bulk modulus for  marcasite type FeS$_2$ is 165.7 GPa from  single crystal elastic constants which is in reasonable agreement with the single crystal bulk modulus.  It is seen that  B$_H$ $>$ G$_H$ for polycrystalline FeS$_2$, which implies that the quantity that limits mechanical stability is G$_H$. Apart from these, we also calculated the Debye temperature ($\Theta$(D)) using sound velocities. $\Theta$(D) is a fundamental quantity that correlates several physical properties such as specific heat, thermal conductivity and melting point of the crystal with  elastic constants. At low temperatures, $\Theta$(D) can be estimated from the average sound velocity ($\upsilon$$_m$), which is the average of longitudinal($\upsilon$$_l$) and transverse($\upsilon$$_t$) sound velocities. The calculated values of $\upsilon$$_l$, $\upsilon$$_t$,  $\upsilon$$_m$ and $\Theta$(D) are shown in Table 2. This is the first qualitative prediction of mechanical properties of marcasite type FeS$_2$.
\begin{table}[ht]
\caption{Single crystal elastic constants (C$_{ij}$, in GPa), Bulk moduli (B$_X$, X=V, R, H ), Shear moduli (G$_X$,  X=V, R, H ) and sound velocities ($\upsilon$$_l$, $\upsilon$$_t$, $\upsilon$$_m$ in km/sec) and Debye temperature (in K) of FeS$_2$. All quantities are calculated at the respective theoretical equilibrium volume obtained with the PBE functional.}
\begin{tabular}{lllllllllll}
\hline
C$_{11}$	&C$_{22}$	&C$_{33}$	&C$_{44}$	&C$_{55}$	&C$_{66}$	&C$_{12}$	&C$_{13}$	&C$_{23}$\\
303.1		&454.3		&322.8		&105.9		&158.2 		&153.9		&47.0		&106.4		&55.8\\
\hline
B$_V$	&B$_R$	&B$_H$	&G$_V$	&G$_R$	&G$_H$	&$\upsilon$$_l$	&$\upsilon$$_t$ &$\upsilon$$_m$	&$\Theta$(D)\\
166.5	&164.8	&165.7	&141.7	&133.9	&137.8	&8.35		&5.25		&5.78		&634.9\\
\hline
\end{tabular}
\end{table}

\subsection{Phonon dispersion and zone centered frequencies}
\par We have studied the phonon dispersion of marcasite  FeS$_2$ as a function of pressure from ambient to 5 GPa using DFPT \cite{dfpt}. Since instability of one or more phonon modes would be indicative of dynamical instability of the structure, we have carried out study of phonon dispersion in the entire first Brillouin zone to investigate the dynamical stability of marcasite structure. The unit cell of marcasite FeS$_2$ contains 6 atoms and hence it has 18 phonon modes for each wavevector, out of which three are acoustic  and remaining 15 modes are optical modes. According to the group theoretical analysis the optical modes at $\Gamma$ - point can be represented as

\begin{equation}
\Gamma= 2A_g+2B_{1g}+B_{2g}+B_{3g}+2A_u+B_{1u}+3B_{2u}+3B_{3u}.
\end{equation}

In this,  A$_u$ mode is inactive, whereas the modes A$_g$, B$_{1g}$, B$_{2g}$, B$_{3g}$ are Raman active and B$_{1u}$, B$_{2u}$, B$_{3u}$ modes are Infrared active. The calculated zone center frequencies at ambient pressure are shown in table 3 accompanied with experimental and other theoretical results. The calculated values are in reasonable agreement with the experiment. In an earlier study, Spangnoli {\it et al} \cite{Spagnoli} reported the zone center frequencies for both pyrite and marcasite structure and their results show  both the structures to have zone centered modes with real frequencies implying  stability of the modes. We also find a similar situation in our case. We have also calculated zone center frequencies as a function of pressure and we observe no imaginary frequencies at zone center. The calculated zone centered vibrational frequencies upto 5 GPa are shown in Fig. 5(a). From this figure it is clear that, there is no softening of zone centered frequencies and it allows us to confirm marcasite to be dynamically stable upto 5 GPa \cite{kanchana}. It is to be emphasized that there are no signatures of the impending pressure induced structural phase transition in the phonon spectra. In addition, the calculated dispersion curves along high symmetry directions and the corresponding phonon density of states for marcasite FeS$_2$ at P=0 GPa and P=4 GPa are shown in Fig. 5(b). There is a considerable overlap between the acoustic and optical modes which can be clearly seen from the phonon dispersion. The optical mode frequencies from 310 cm$^{-1}$ to 390 cm$^{-1}$ are dominated by S-atoms. We do not find any imaginary phonon frequencies in the phonon dispersion curves along any direction of the Brillouin zone at ambient and high pressure. This clearly establishes the dynamical stability of marcasite FeS$_2$. However, since linear response theory is based on harmonic approximation, anharmonic effects are ignored in the present calculations. It is plausible that anharmonic effects may also play a role in driving the observed structural phase transition. We have also calculated Raman and IR spectra of marcasite FeS$_2$ from ambient to 5 GPa. From this, we found that Raman peak intensities decrease with increasing pressure, whereas IR peak intensities increase with increasing pressure. However, the peaks are shifted to higher frequencies in both the cases. The calculated IR and Raman spectra at ambient and 5 GPa are shown in Fig. 5(c), (d).

\begin{table}[ht]
\caption{Comparison of present calculated Phonon frequencies with experimental and other theoretical vibrational frequencies (cm$^{-1}$) of marcasite FeS$_2$ at 0 GPa.}
\begin{tabular}{lllllllll}
\hline
&Mode		&	&Present work		&Exp$^e$	&Theory$^f$\\
\hline\\
&A$_u$		&	&202.64			&inactive	&207 \\

&B$_{2u}$	&	&248.82			&325		&248 \\

&B$_{3u}$	&	&288.64			&293		&279\\

&A$_g$		&	&305.23			&323		&317\\

&B$_{2u}$	&	&317.95			&404		&323 \\

&B$_{3g}$	&	&320.20			&367		&339\\

&B$_{2g}$	&	&322.60			&308		&342 \\

&B$_{3u}$	&	&361.90			&353		&360\\

&A$_u$		&	&366.57\\	

&B$_{3u}$	&	&367.76			&387		&373 \\	

&A$_{g}$	&	&377.43			&386		&388\\

&B$_{1g}$	&	&379.25			&377		&382\\

&B$_{2u}$	&	&392.13			&432		&385\\

&B$_{1u}$	&	&399.79			&404		&409 \\

&B$_{1g}$	&	&456.56 				   &455\\
\hline
$e:$ Ref.\citenum{sourisseau}\\
$f:$ Ref.\citenum{Spagnoli}
\end{tabular}
\end{table}

\subsection{Transition from marcasite to pyrite}
\par On the basis of the total energy calculations of marcasite and pyrite FeS$_2$ shown in Fig. 2, we infer possibility of a structural phase transition from the ground state marcasite to pyrite structure. It is also to be mentioned that earlier experiments found pyrite to exist only at high pressures \cite{Ennaoui,Merkel}. Interestingly, we did not observe any phonon softening or structural variations in marcasite FeS$_2$ by our calculations under pressure.  Generally, most of the MX$_2$ type compounds crystallize in pyrite, marcasite or arsenopyrite structures. These structures are closely related to each other  \cite{Hulliger}. Marcasite structure is again classified into {\it regular marcasite} and {\it anomalous marcasite} depending on the c/a ratio and bond angle (M-X-M) between the neighboring cations in the edge shared octahedra. A c/a ratio around 0.53-0.57 and bond angle is less than 90$^\circ$ correspond to regular marcasite, whereas anomalous marcasite has c/a ratio of 0.73-0.75 and bond angle is greater than 90$^\circ$. In the present study, the calculated c/a ratio and bond angle is found to be 0.76 and 98.6$^\circ$  respectively for marcasite FeS$_2$. From this it is clear that FeS$_2$ belongs to the class of anomalous marcasite.

\par We now proceed to describe the mechanism that drives marcasite to pyrite transition under hydrostatic pressure. The similarities and differences between the two competing structures are quite evident when the marcasite supercell spanned by lattice translation vectors $\vec{a}'$ = $\vec{a}+\vec{c}$, $\vec{b}'$ = $-\vec{a}+\vec{c}$, and $\vec{c}'$ = $\vec{b}$ is compared with the conventional unit cell of pyrite- see Fig. 1(a-c). Here, $\vec{a}$, $\vec{b}$ and $\vec{c}$ are the lattice translation vectors of the conventional unit cell of the marcasite structure.  Both the cells contain the same number of atoms.  The Fe atoms and the centers of the S$_2$ dimers have the same fractional coordinates  in both the cells. While all the faces of the pyrite unit cell are squares, the $\vec{a}'-\vec{c}'$ and $\vec{b}'-\vec{c}'$ faces the marcasite supercell are almost squares, whereas the $\vec{a}'-\vec{b}'$ face of the marcasite is a rhombus with the angle 105.37$^{\circ}$ between $\vec{a}'$ and $\vec{b}'$. Furthermore, while the orientation of the S$_2$ dimers with centers lying on lines parallel to the $\vec{a}'$ direction of marcasite cell are the same, it flips by $\pi/2$ along alternate lines in the pyrite structure. In the present calculation we found that, the marcasite structure is most compressible along $\vec{a}$, less compressible along $\vec{c}$ and least compressible along $\vec{b}$-direction. Therefore when a sufficiently large hydrostatic pressure is applied, there is a differential compression along $\vec{a}$ as compared to the $\vec{c}$ direction of marcasite cell. This results in reducing the angle between $\vec{a}'$ and $\vec{b}'$. When the pressure is increased further, this differential change can trigger a flipping of the S$_2$ dimers along alternate lines parallel to $\vec{a}'$ direction by $\pi/2$ thereby facilitating a discontinuous transformation of the rhombus into a square. This results, together with relaxation of atomic positions leads to the marcasite-pyrite transformation.

\section{Electronic and thermoelectric properties}
\subsection {Band structure and density of states}
\par Quite a good number of electronic structure calculations are reported in the literature aimed at understanding the band structure and density of states of both the marcasite and pyrite structures. It is well known that the thermoelectric properties are quite sensitive to the details of the band structure. It is therefore clear that the reliability of the computed thermoelectric properties would depend on the accuracy of the electronic structure calculations. In this perspective, we have repeated the band structure and density of state calculations using TB- mBJ potential, which is well known to reproduce accurately the experimental band gap values \cite{Tran,Koller1,Koller2,Dixit,Hong}. The calculated band structure for both orthorhombic marcasite at 0 GPa and cubic pyrite at 4 GPa along the high symmetry directions of the Brillouin zones (see Fig. 6(a) and (b)) are shown in Fig. 7(a) and 7(b). From the band structure analysis, we have seen that the conduction band minimum (CBM) and valence band maximum (VBM) are located at two different high symmetry points in the Brillouin zone making the material an indirect band gap semiconductor in both the structures. The calculated band gap of marcasite is found to be 1.603 eV and in pyrite it is 1.186 eV. In the case of marcasite, we predict band gap to be much higher than the experimental band gap of 0.34 eV \cite{Jagadesh} obtained from resistivity measurement. It must be mentioned that this observation is similar to the results obtained using other exchange correlation functions \cite{Ruoshi}. In the case of pyrite, there is a wide range of band gaps from 0.7-2.62 eV reported earlier \cite{Ennaoui,Ferrer,Bullett,Opahle,Will,Ahrens}, and this spread in values of the energy gap may be due to the
experimental limitations as mentioned by Ennaoui et al. and Ferrer  et al \cite{Ferrer,Ennaoui}. The photo conductivity measurements show a consistent band gap for the pyrite in the range of 0.9-0.95 eV which is in good agreement with the optical and conductivity experiments \cite{Ennaoui,schlegel}, and also in good agreement with our present calculations and the recent calculations by Jun Hu it et al \cite{JunHu}. We also investigated dependence of the value of band gap on the position of S atoms in the marcasite unit cell. Unlike the extreme sensitivity of E$_g$ seen in the earlier calculations for the pyrite structure\cite{Eyert}, we did not find any significant variation in the band gap of marcasite with position of S atoms.  From the band structure analysis, we find that the CBM is located between $\Gamma$ and $X$ points in the Brillouin zone, whereas in the case of pyrite, we find it at the center of the Brillouin zone. The VBM in the case of marcasite is located along the $\Gamma$ - $Y$ direction with highly dispersed band, whereas in case of pyrite it is clearly seen at $X$ point with less dispersion towards the $\Gamma$ point. These bands mainly arise from the Fe-d and S-p states in both the structures, but in the case of marcasite, the
contribution of these states are low compared to the pyrite. The principal aim of the present work is to calculate the thermoelectric properties of FeS$_2$ and its variation with carrier concentration. It is necessary to estimate the effective masses of the carriers in various electron and hole pockets to achieve this task.  We have calculated the mean effective mass of the carriers at the conduction and valence band edges by fitting the energy of the respective bands to a quadratic polynomial in the reciprocal lattice vector $\vec{k}$. The calculated effective masses for both marcasite and pyrite structures in some selective directions of the Brillouin zone are tabulated in Table 4. It is quite clear that the bands are less dispersive in the pyrite structure almost in all the high symmetry directions. This would imply large effective mass for the carriers belonging to these bands and hence a high thermopower. However, presence of carriers with large  mobility is required for obtaining a higher electrical conductivity. Thus there is a possibility of obtaining large ZT factor in materials possessing multiple pockets of carriers with large and small effective masses with the former one leading to large S, and the latter one enhancing $\sigma$\cite{DJS,aggate2}.  It is interesting to note that the electronic structure of both the phases of FeS$_2$ reveals presence of multiple carrier pockets with substantially different effective masses thereby suggesting that they may be having good thermoelectric properties.

\begin{table}[ht]
\caption{The calculated effective mass of the marcasite and pyrite in some selective direction of the Brillouin zone in the units of electron rest mass.}
\begin{tabular}{llllllll}
\hline
		&Direction	&&Valance band	&&Conduction band\\
\hline
Marcasite	&$\Gamma$ - X	&&0.118		&&0.490	\\
		&$\Gamma$ - Y	&&0.081		&&0.058 \\	
		&$\Gamma$ - Z	&&0.062		&&0.099 \\
\hline
Pyrite		&$\Gamma$ - X	&&0.512		&&0.019	\\
		&$\Gamma$ - M	&&0.656		&&0.116 \\	
		&$\Gamma$ - R	&&0.096		&&0.033 \\
\hline
\end{tabular}
\end{table}

\par The calculated Density of States (DOS) along with the l-projected DOS for both the structures are shown in Fig. 8(a) and 8(b). In Fig. 8(b), it is clearly seen that there is a sudden increase in DOS at the VBM (E$_F$), and this is also quite evident from the low dispersion seen in the band structure (Fig. 7(b)). In the case of marcasite (Fig. 8(a)) we find an increase in the DOS of valence band upto -0.2 eV, then it dips a little near -0.2 eV, and then increases  again. We find a similar increasing trend in the DOS in case of conduction band for both the structures with a small variation in the marcasite. It is also evident from the above band structure and DOS that p-type doping is more favorable for obtaining better thermoelectric properties than n-type doping. Optimized doping level and transport properties are discussed in the succeeding section.

\subsection{Thermoelectric properties}
The recent experimental study of Lu {\it et al} \cite{xulu} reveals that natural minerals are good candidate materials for achieving high thermoelectric efficiency. In view of this, we have attempted to calculate the thermoelectric properties of both marcasite and pyrite phases of FeS$_2$. Towards this end, we have calculated the thermopower and electrical conductivity/scattering time ( $\sigma$/$\tau$), as a function of carrier concentration and temperature, using the Boltzmann transport equation approach as implemented in BOLTZTRAP\cite{Madsen} code. These properties were calculated  at ambient pressure for the marcasite phase and at high pressure for the pyrite phase. The results are discussed in the following sections.

\par The thermopower plays a vital role in deciding the performance of the thermoelectric material, the reason being the direct proportionality of the figure of merit to the square of the thermopower (ZT=S$^2$$\sigma$T/$\kappa$). The thermopower calculated using the electronic structure data  depends on the carrier  concentration and temperature. In view of the constant scattering time approximation that is employed, energy dependence of $\tau$ is ignored. The calculated thermopower as a function of hole and electron concentrations at different temperatures are shown in Fig. 9(a) and 9(b). Here we have given the results for 300 K, 400 K and 500 K for marcasite and 700 K, 800 K and 900 K for the pyrite, as marcasite is reported to be stable only till 573 K (300 $^\circ$C), found to exist in mixed state between 573-673 K (300-400 $^\circ$C) and completely transformed to pyrite above 673 K (400 $^\circ$C) \cite{Kjekshus,Fleet,Gronvold,Murowchick,Lennie}. We observe the thermopower to increase with decreasing carrier concentration for both electron and hole doping which implies absence of bipolar conduction in the optimum concentration region. At low carrier concentrations (hole concentration of $1\times 10^{19}$ cm$^{-3}$ at 900 K for pyrite), we can see a decrease in the thermopower which indicates onset of bipolar conduction at those regions. We also found the Pisarenko behavior ({\it i.e.} logarithmic variation) in the thermopower, in the range of 10$^{19}$-10$^{21}$ cm$^{-3}$ which is an optimum working region for any good thermoelectric material. In this region, we find the thermopower of marcasite with a hole concentration of 1x10$^{19}$ cm$^{-3}$ to vary  between 550-610 $\mu$ V/K for the  temperature range 300 K-500 K. In case of the pyrite, the thermopower is higher $\sim 750 \mu $V/K up to temperature $\sim$ 800 K, followed by a reduction at $\sim$ 900 K due to the bipolar conduction.  The difference between the thermopower values of marcasite and pyrite may be due to difference in effective masses of the carriers near the Fermi level of both the structures (see Fig. 7(a) and 7(b)). Hole doping leads to larger thermopower as compared to electron doping in both the structures. The computed range of thermopower is in good agreement with the earlier experimental investigation of Maria Telkes \cite{Maria} for both the structures. This indicates that the marcasite and pyrite phases of FeS$_2$ are good thermoelectric materials with both hole and electron doping, but hole doping is preferred over electron doping. This is also evident from the Fig. 8(b) as discussed earlier. Apart from this, we can also see that marcasite is favourable for the low temperature(upto 500 K) thermoelectric applications and pyrite is favorable for the high temperature(upto 900 K) thermoelectric applications.

\par The other important factor which influences the thermoelectric figure of merit is the electrical conductivity $\sigma$. While $\sigma$/$\tau$ is an intrinsic property of any material, the relaxation time $\tau$ also depends on the nuances of material preparation. Therefore, we first consider variation of  $\sigma$/$\tau$ with carrier concentration and temperature. Results of our calculation of  $\sigma/\tau$, as a function of both the electron and hole concentrations for both the structures are shown in Fig. 10(a) and 10(b). We found that there is no significant change in $\sigma/\tau$ with temperature in the temperature range of our interest when the carrier concentration is varied in the range $1\times 10^{19} - 1\times 10^{21}$  for both the structures. We also observe that $\sigma/\tau$ varies from $1\times 10^{18}$-$3\times 10^{19}$ {\it $(\Omega ms)^{-1}$} for marcasite and from $6\times 10^{17}$-$3\times 10^{19}$ {\it $(\Omega ms)^{-1}$} for the pyrite for the optimum hole concentration at the investigated temperatures. Since no general method exists to calculate microstructure dependent relaxation times of electrons, we are not in a position calculate $\sigma(T)$  from first principles. However, we have attempted to estimate the relaxation times of electrons at a fixed temperature and carrier concentration of the pyrite phase from the limited experimental data available on naturally occurring pyrite\cite{kato}. We have then used the same relaxation time for the marcasite structure as well, since these two polymorphic phases are closely related. For the naturally occurring pyrite with an electron concentration of 3.3$\times$ 10$^{18}$ cm$^{-3}$, the resistivity and thermal conductivity of natural pyrite was found to be of the order of 2.17$\times$ 10$^{3}$ {\it $\Omega^{-1}$ m$^{-1}$} and 24.89 {\it W/m K} respectively at a temperature of 578 {\it K}\cite{kato}. The corresponding  relaxation time $\tau$ is estimated to be 101.1 $\times$ 10$^{-14}$so that the computed resistivity matches the experimental value. Using this relaxation time, $\sigma$ is calculated for various carrier concentrations. The value of ZT thus computed is found to be 0.32 for a concentration 3 $\times$ 10$^{19}$ at a temperature of 700K. Similarly we found a ZT value of 0.14 for marcasite at a temperature of 300 {\it K}. The value of ZT can be increased further by reducing the thermal conductivity by resorting to nano-structuring technique\cite{nature,science} as well as by improving $\sigma$ by using phase pure materials. Our theoretical calculations give the guidelines for further  experimental investigation in this regard.

\section{Conclusion}
In summary, we have reported a theoretical description of the structural transition of natural mineral FeS$_2$. From this study, we conclude that  the ground state of  FeS$_2$ is the marcasite structure under ambient conditions, and it transforms to the pyrite structure at high pressures. The calculated structural properties such as lattice parameters, internal coordinates of sulphur atoms, bond lengths and bulk modulus at ambient conditions are in good agreement with available experiments and other theoretical reports.  We have also predicted the single crystal and polycrystalline elastic properties and confirm that marcasite FeS$_2$ is mechanically stable at ambient conditions. Our calculations using pressure coefficients of three lattice parameters and single crystal elastic constants confirmed that the marcasite phase is least compressible along the $\vec{b}$- axis.   Furthermore, the dynamical stability of marcasite FeS$_2$ is studied by phonon dispersion and the calculated zone centered frequencies at ambient and high pressure are in good agreement with the earlier experiments and theoretical results. It is interesting to note that the calculated zone centered frequencies and the phonon dispersion up to 5 GPa along different directions do not show any softening under pressure, reflecting the dynamical stability of marcasite FeS$_2$ up to 5 GPa. A discussion of marcasite to pyrite structural transition is presented based on a detailed comparison of the geometry and energetics of the two structures. We have also calculated the electronic band gap using semi-local exchange correlation functional TB-mBJ and found that both polymorphic structures are indirect band gap semiconductors. Finally, we have calculated thermoelectric properties and find that the thermopower for the high pressure phase is relatively higher compared to the ambient phase. We also predict that marcasite can be used for low temperature thermoelectric applications, whereas pyrite can be used for the high temperature applications. We hope that our work on the transport properties will further stimulate the experimentalists for a detailed study of the thermoelectric properties of this interesting mineral.

\begin{acknowledgement}
Authors thank Dr. G. Parthasarathy, CSIR-NGRI for bringing interest towards the structural stability of natural marcasite. V. K. G and V. K acknowledge IIT-Hyderabad for the computational facility. Authors  S. A and G. V thank  Center for Modelling Simulation and Design-University of Hyderabad (CMSD-UoH) for providing computational facility. V. K. thank NSFC awarded Research Fellowship for International Young Scientists under Grant No. 11250110051. \\
\end{acknowledgement}

{}

\begin{figure}
\begin{center}
\subfigure[]{\includegraphics[height = 2in, clip]{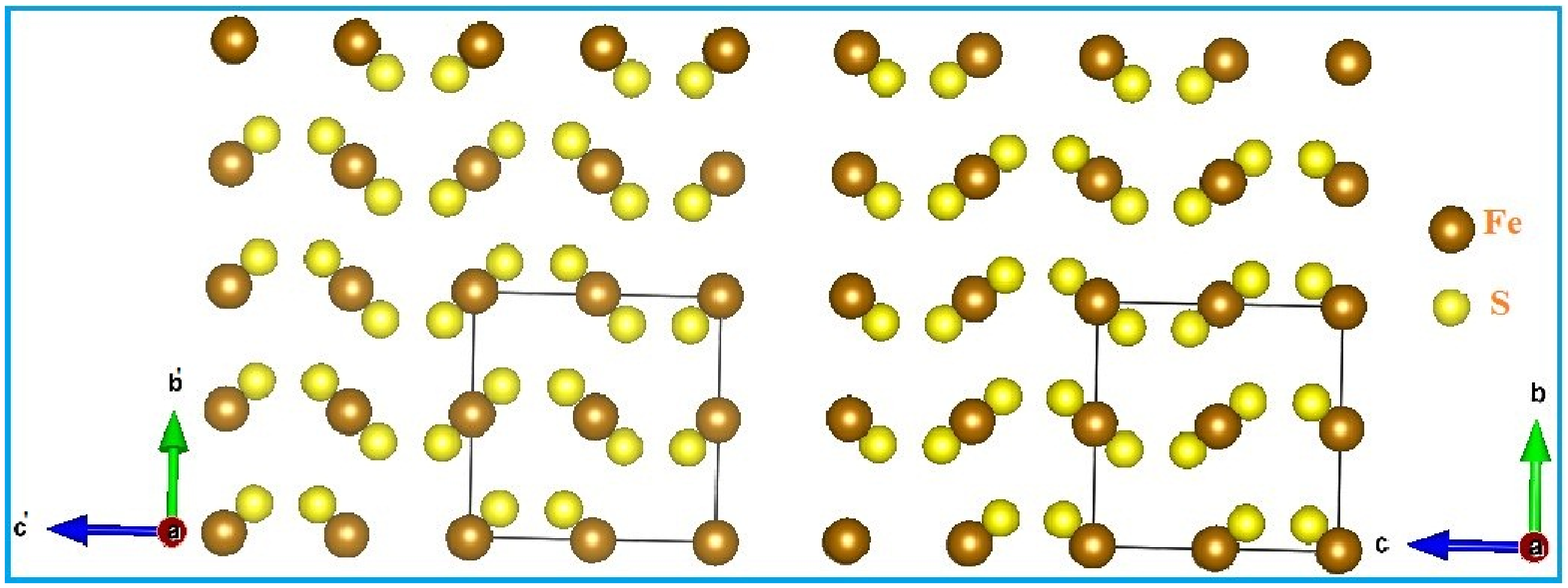}}\\
\subfigure[]{\includegraphics[height = 2in, clip]{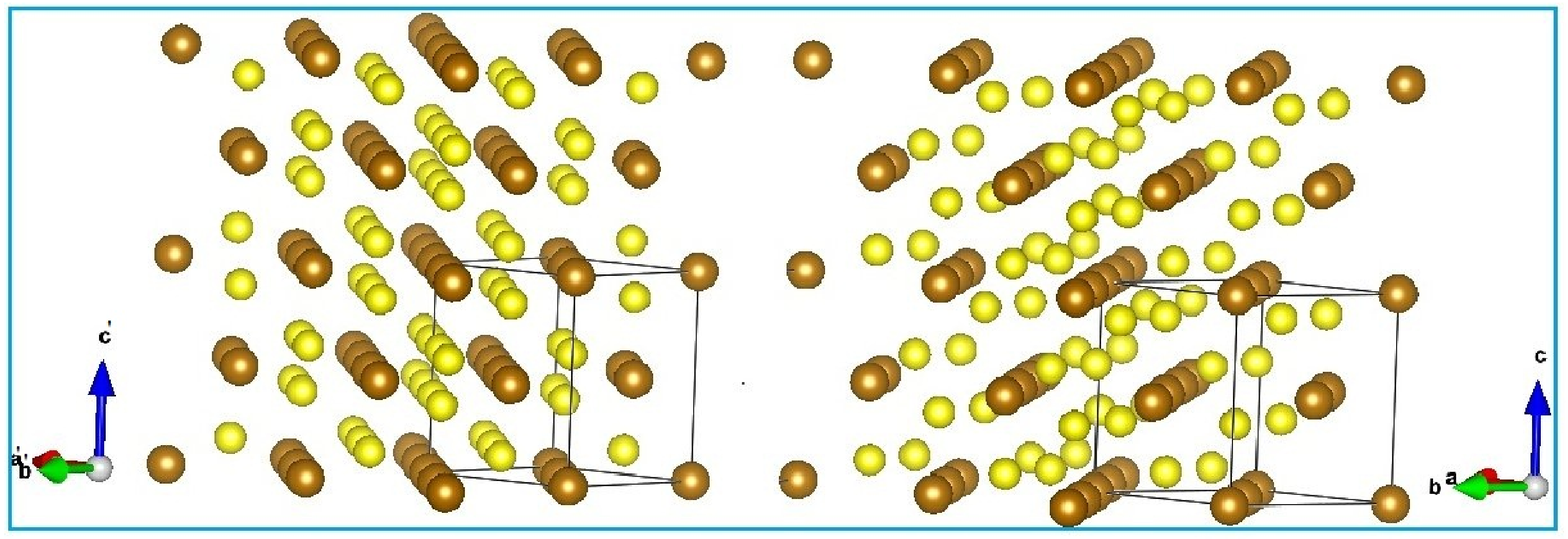}}\\
\subfigure[]{\includegraphics[height = 2in, clip]{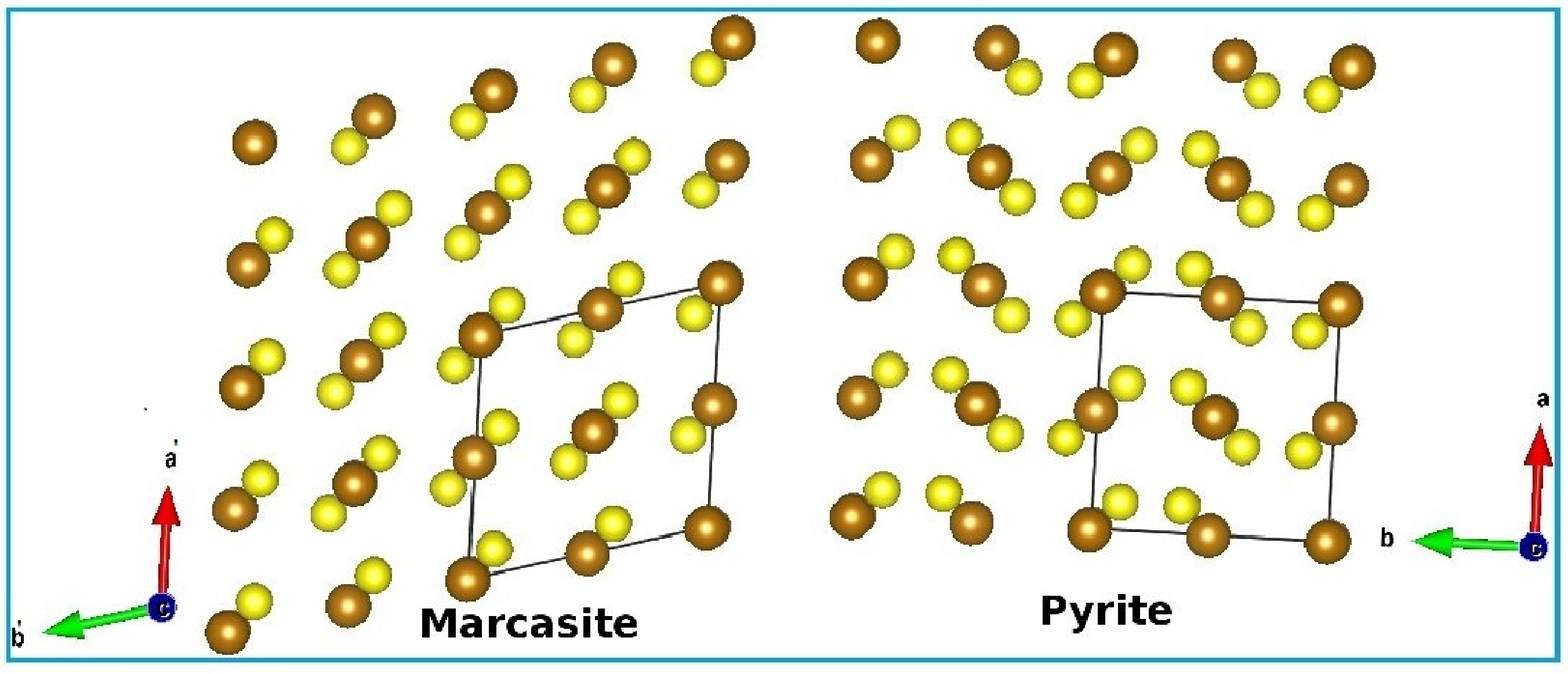}}
\caption{(Color online) The marcasite supercell spanned by lattice translation vectors $\vec{a}'$ = $\vec{a}+\vec{c}$, $\vec{b}'$ = $-\vec{a}+\vec{c}$, and $\vec{c}'$ = $\vec{b}$ is compared with the conventional unit cell of pyrite. Here, $\vec{a}$, $\vec{b}$ and $\vec{c}$ are the lattice translation vectors of the conventional unit cell of the marcasite structure. The (a) $\vec{b}'-\vec{c}'$ and (b) $\vec{a}'-\vec{c}'$ faces the marcasite supercell are almost squares, whereas the (c) $\vec{a}'-\vec{b}'$ face of marcasite is a rhombus.}
\end{center}
\end{figure}

\begin{figure*}[ht]
\begin{center}
{\includegraphics[clip, height = 3.2in]{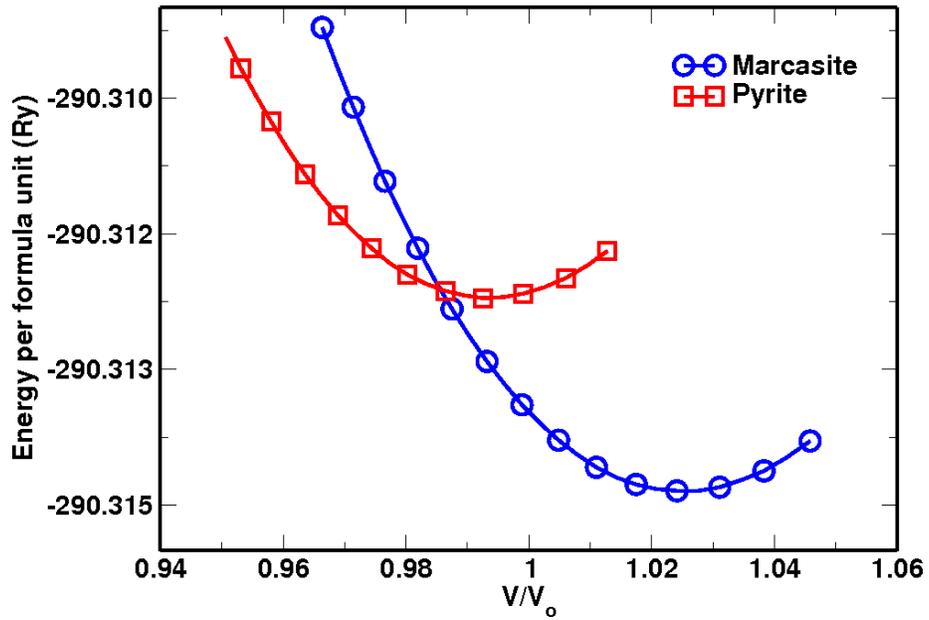}}
\caption{(Color online) Total energy curves as a function of relative volume for FeS$_2$; circles and squares represents marcasite and pyrite, respectively.}
\end{center}
\end{figure*}.

\begin{figure}[ht]
\begin{center}
\subfigure[]{\includegraphics[height = 2.2in, width=3.0in]{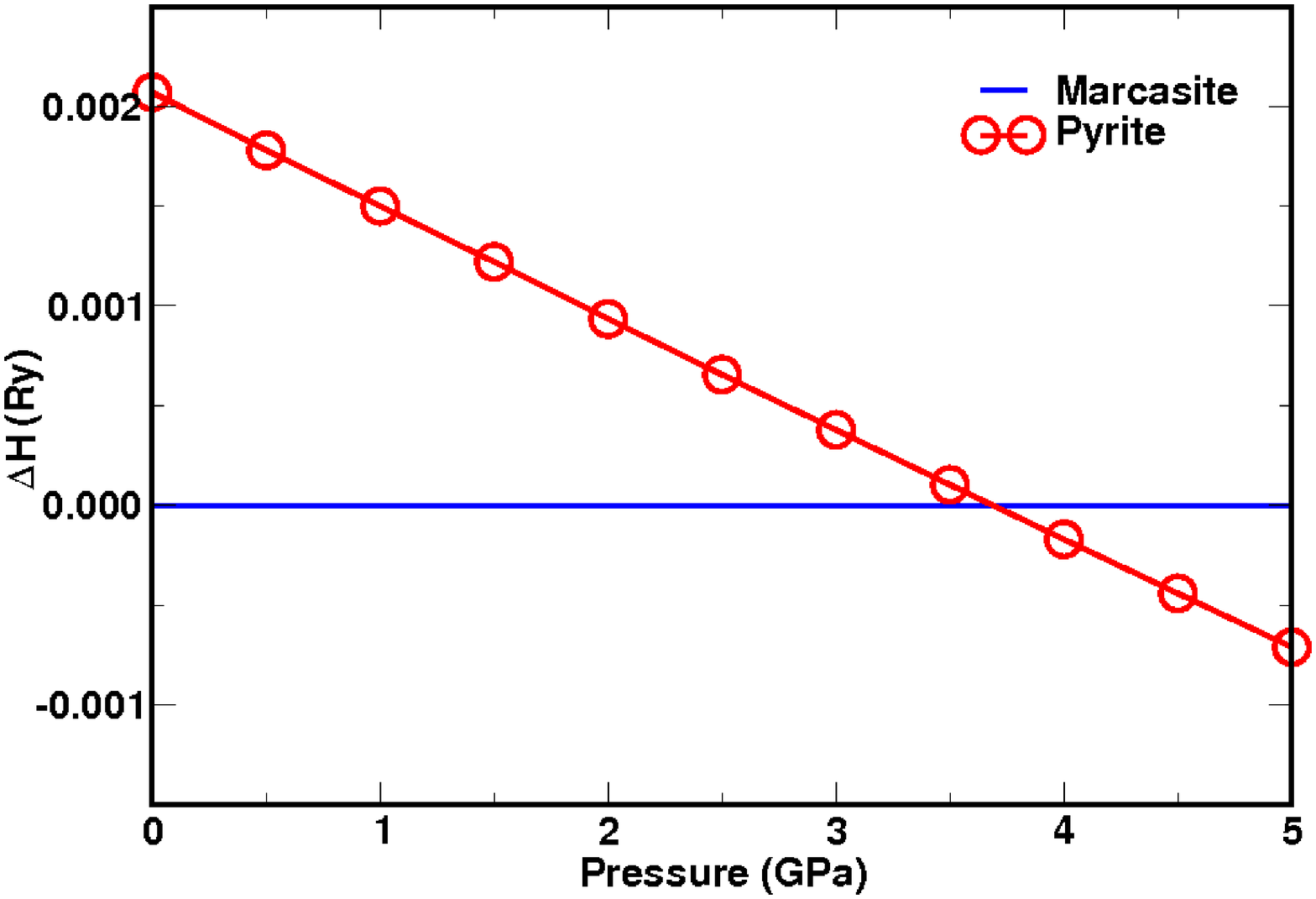}}
\subfigure[]{\includegraphics[height = 2.2in, width=3.0in]{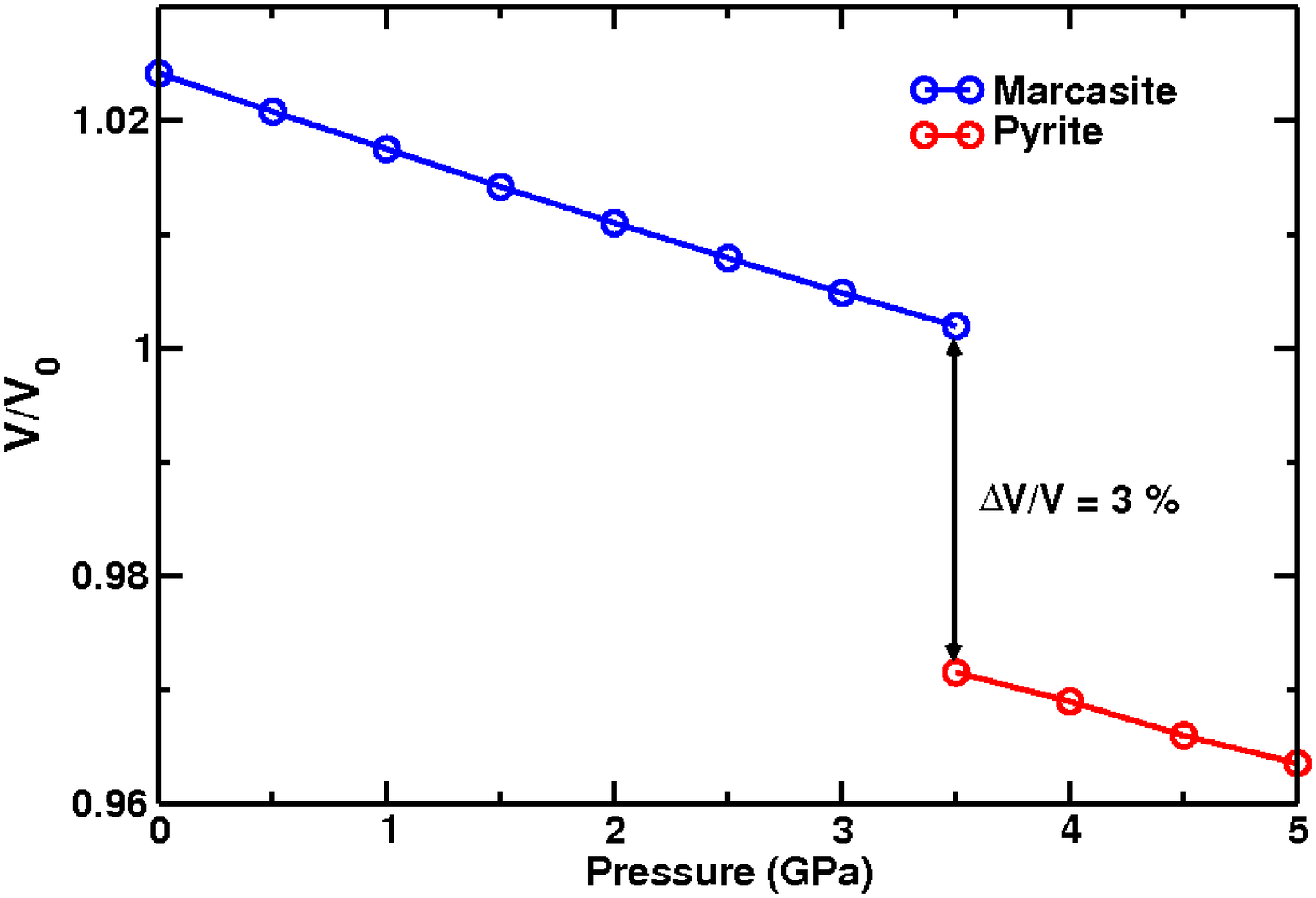}}
\caption{(Color online) Phase transition from marcasite to pyrite per unit cell. (a) Change in enthalpy and (b) Corresponding volume collapse with respective to the pressure.}
\end{center}
\end{figure}

\begin{figure}[ht]
\begin{center}
\subfigure[]{\includegraphics[height = 2.2in,width=3.0in]{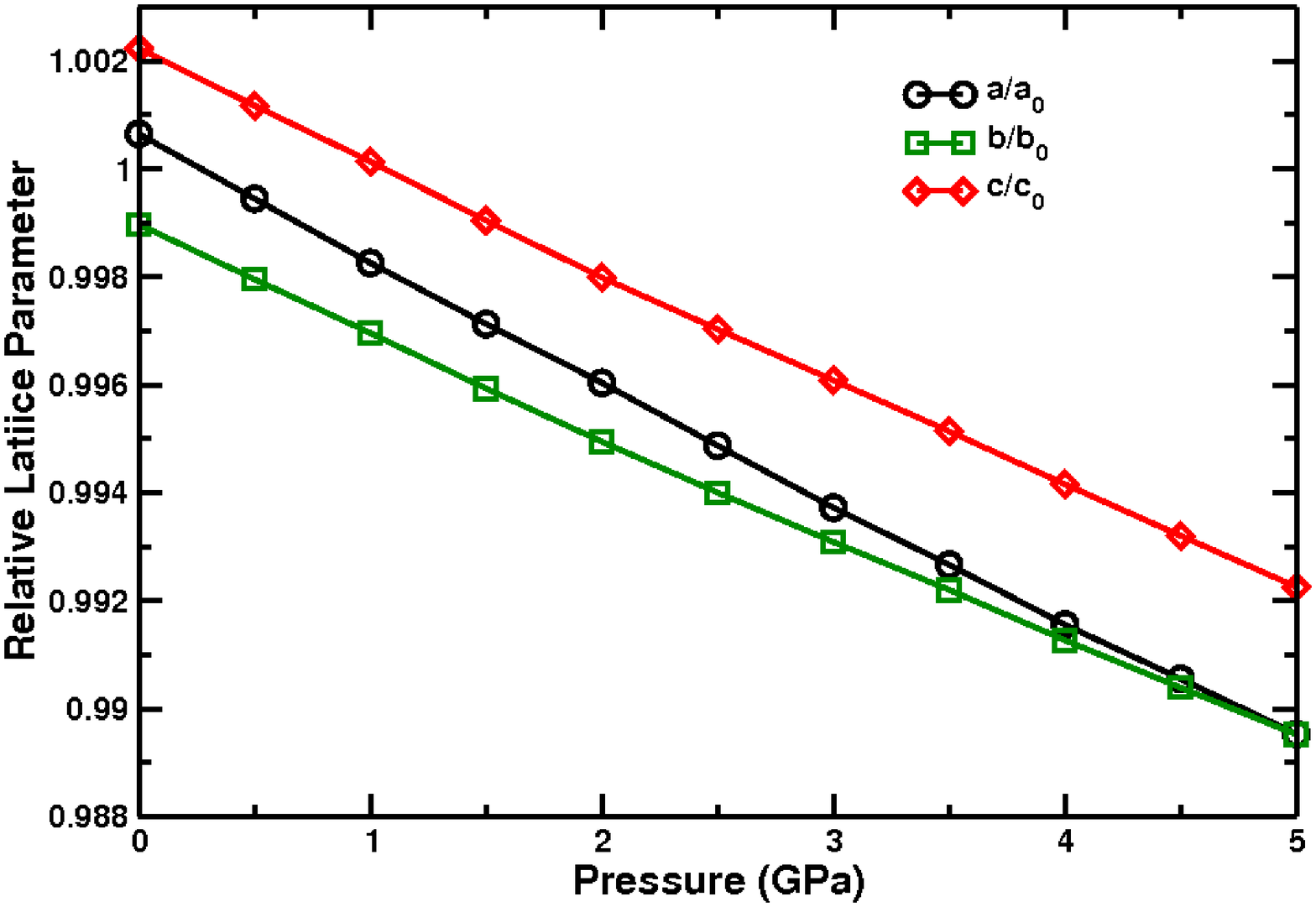}}
\subfigure[]{\includegraphics[height = 2.2in,width=3.0in]{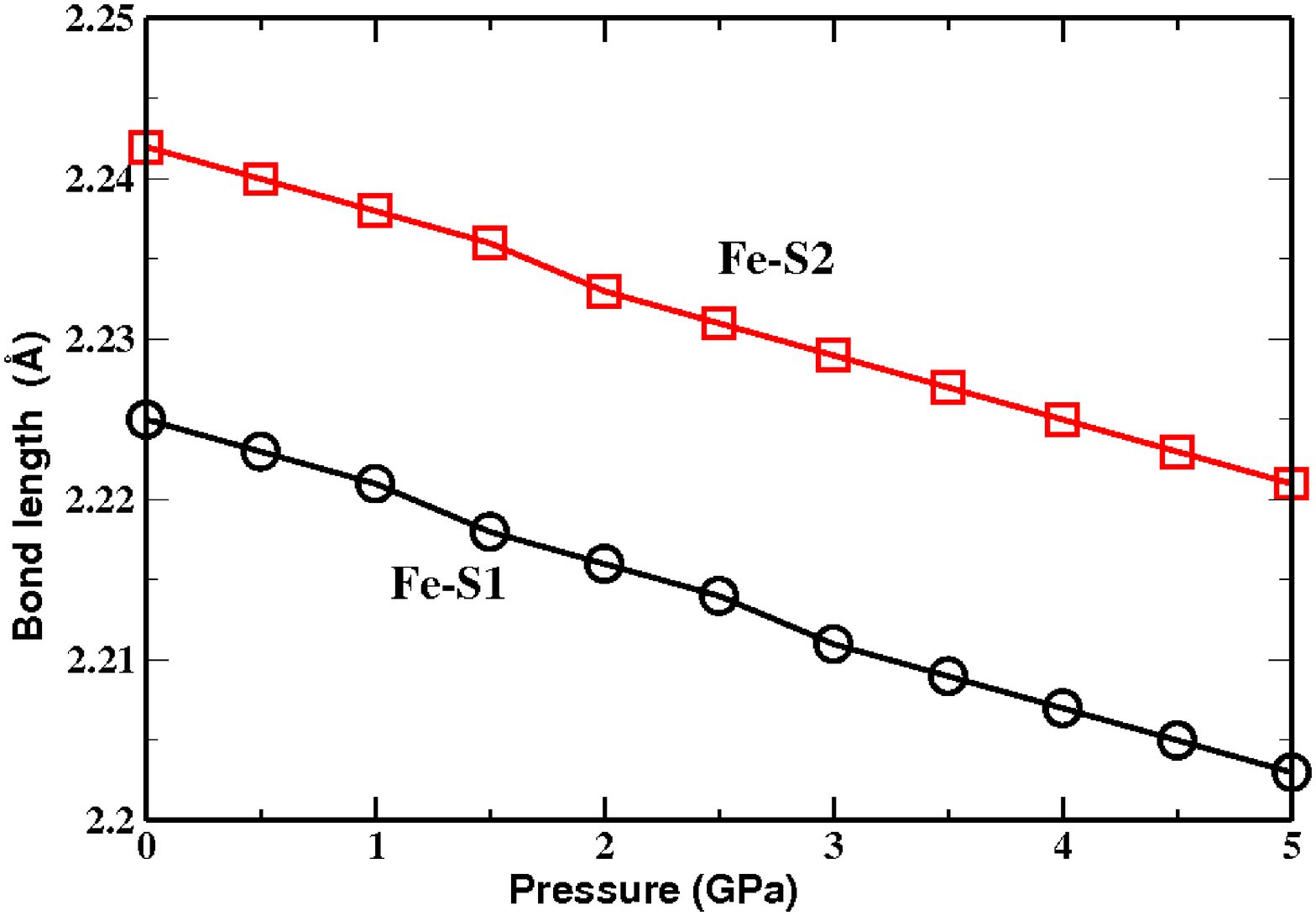}}
\caption{(Color online) Evolution of structural properties of marcasite with pressure up to 5 GPa. (a) Relative lattice parameters and (b) Bond lengths.}
\end{center}
\end{figure}

\begin{figure}[ht]
\begin{center}
\subfigure[]{\includegraphics[height = 2.3in,width=3.0in]{9}}
\subfigure[]{\includegraphics[height = 2.2in,width=3.0in]{10}}
\subfigure[]{\includegraphics[height = 2.2in,width=3.0in]{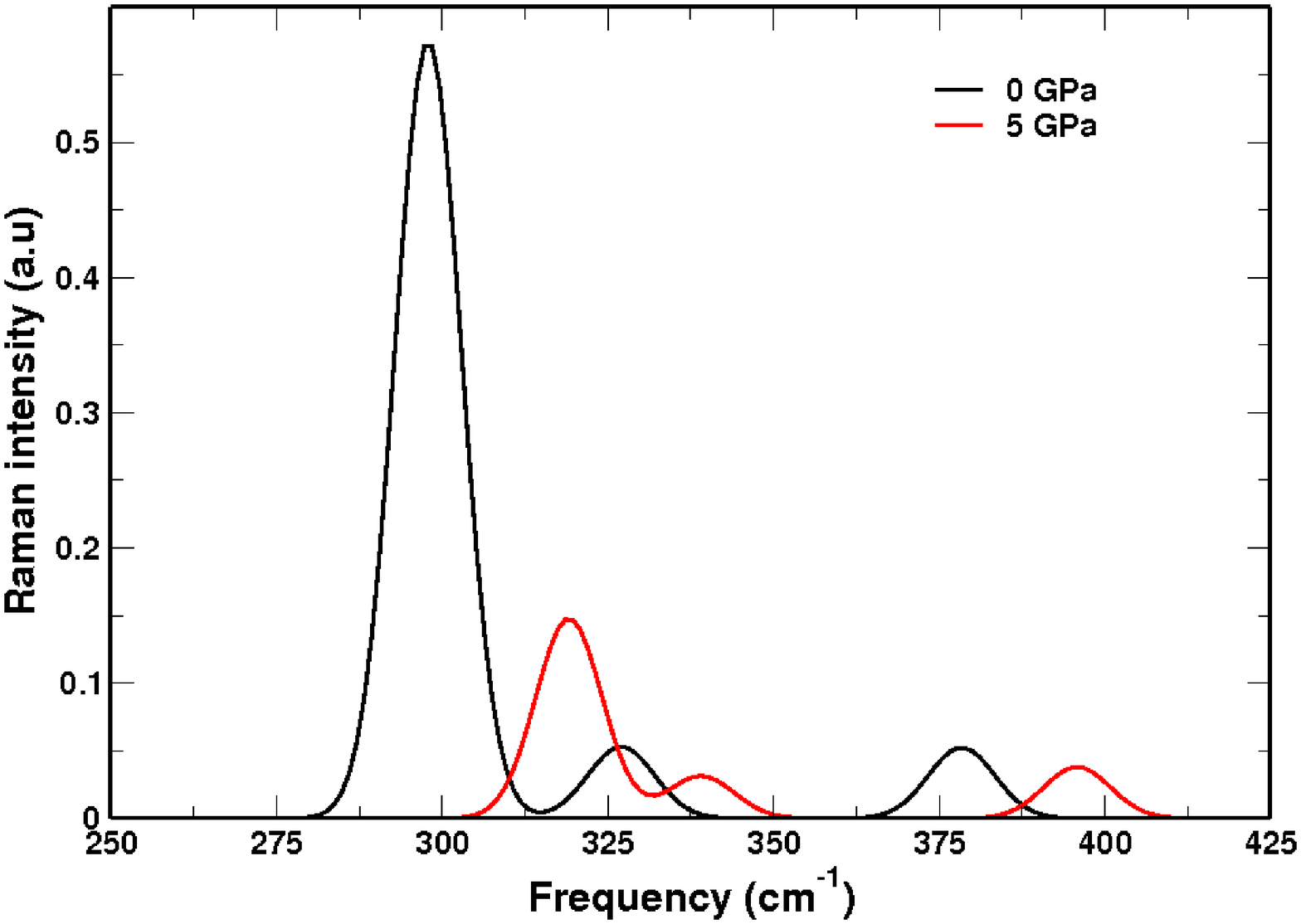}}
\subfigure[]{\includegraphics[height = 2.3in,width=3.0in]{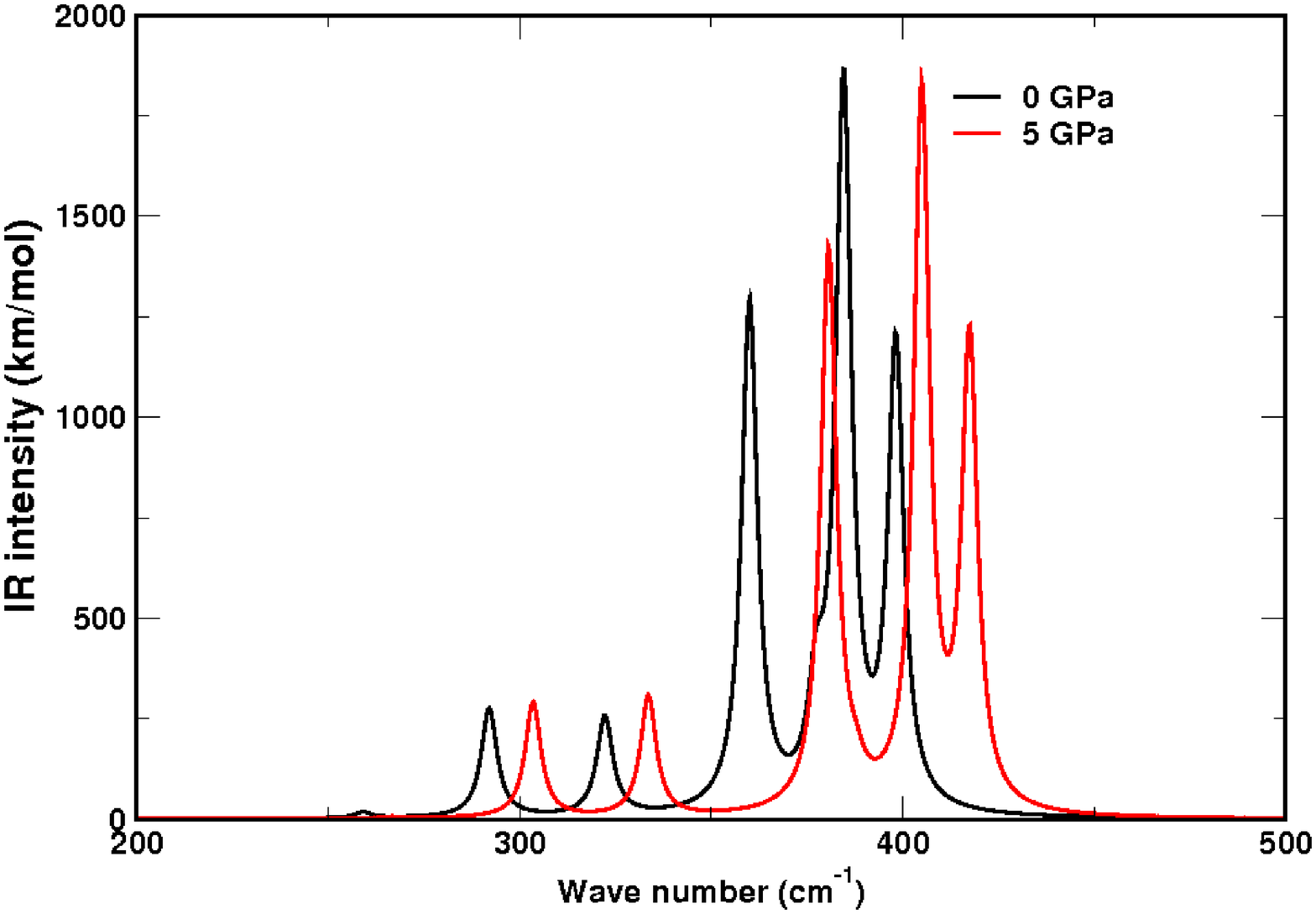}}
\caption{(Color online) Vibrational properties of marcasite FeS$_2$, (a) Zone centered vibrational frequencies from 0 to 5 GPa  (b) Phonon dispersion along high symmetry directions and total phonon density of states at 0 GPa and 4 GPa (c) Raman spectra at 0 GPa and 5 GPa and (d) IR spectra at 0 GPa and 5 GPa.}
\end{center}
\end{figure}

\begin{figure}
\subfigure[]{\includegraphics[width=60mm,height=60mm]{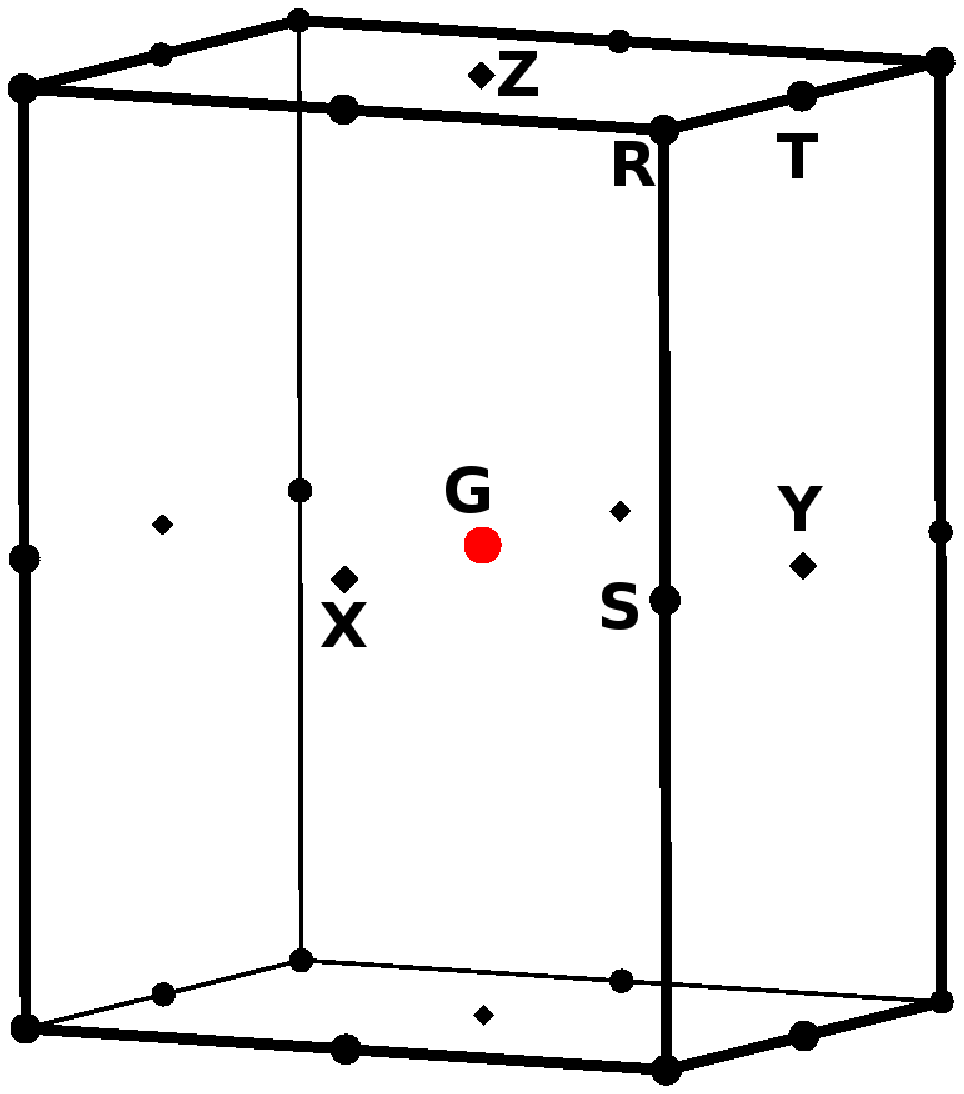}}
\subfigure[]{\includegraphics[width=60mm,height=60mm]{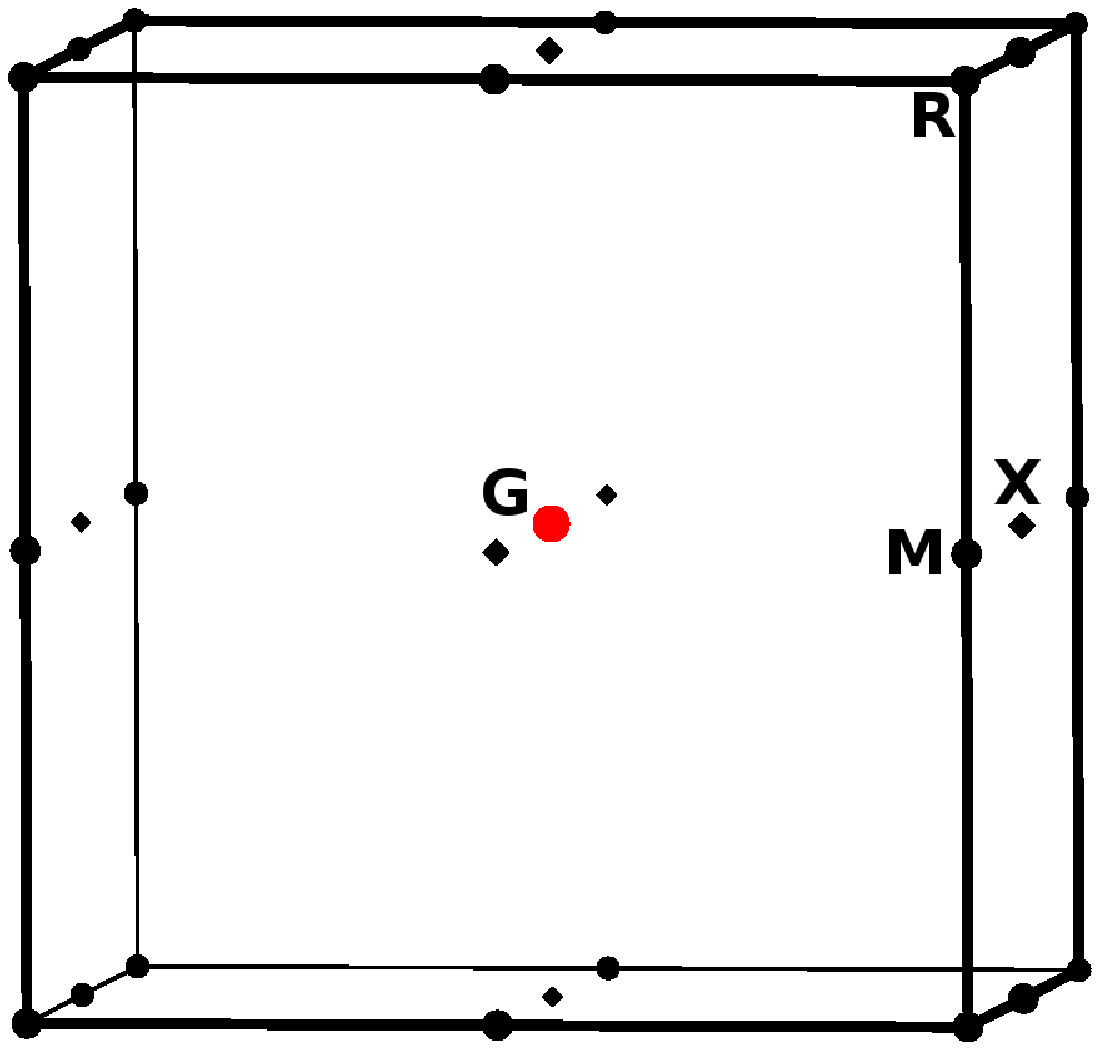}}
\caption{(Color online) Brillouin zone of (a) Marcasite and (b) Pyrite.}
\end{figure}

\begin{figure}
\subfigure[]{\includegraphics[width=75mm,height=75mm]{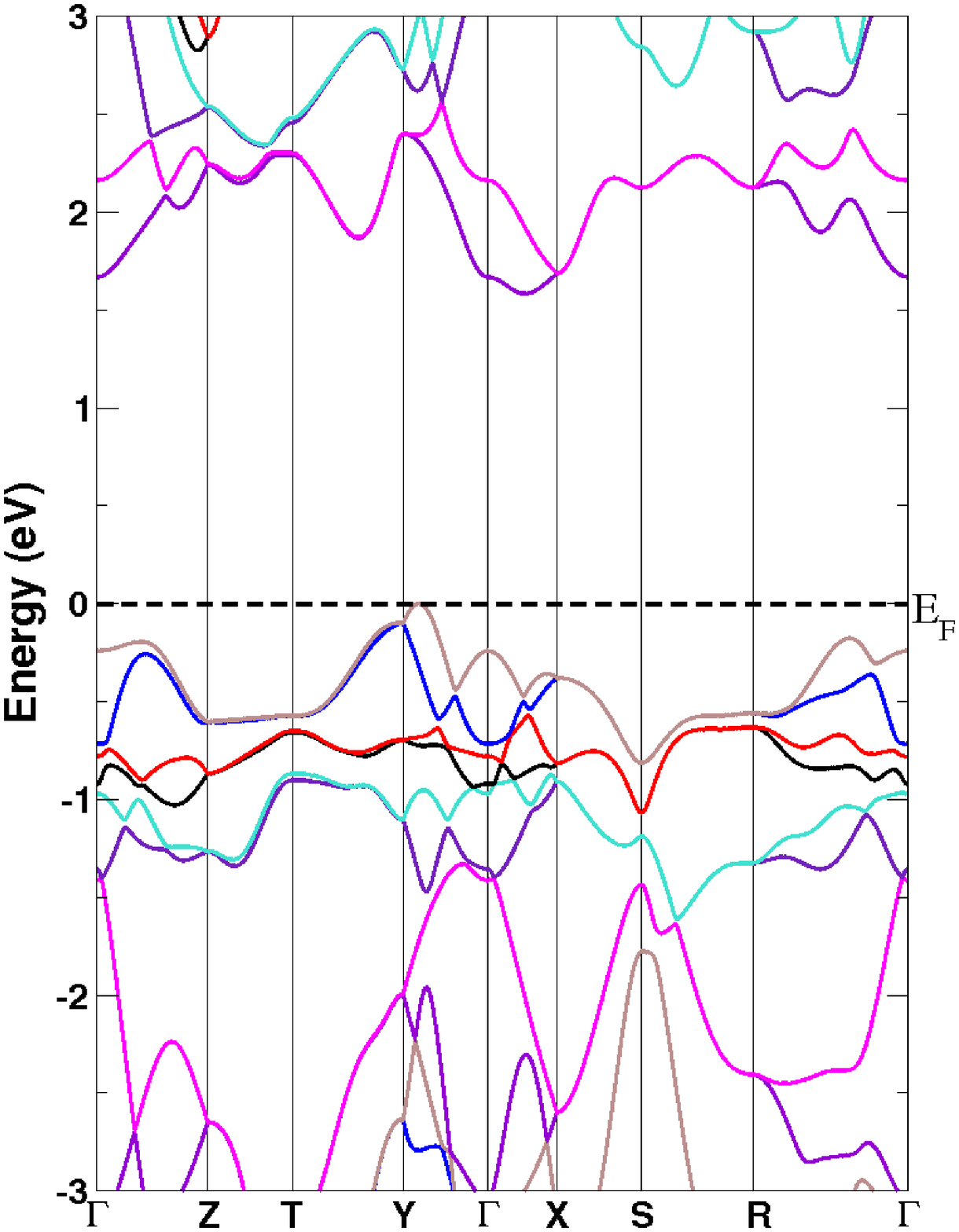}}
\subfigure[]{\includegraphics[width=75mm,height=75mm]{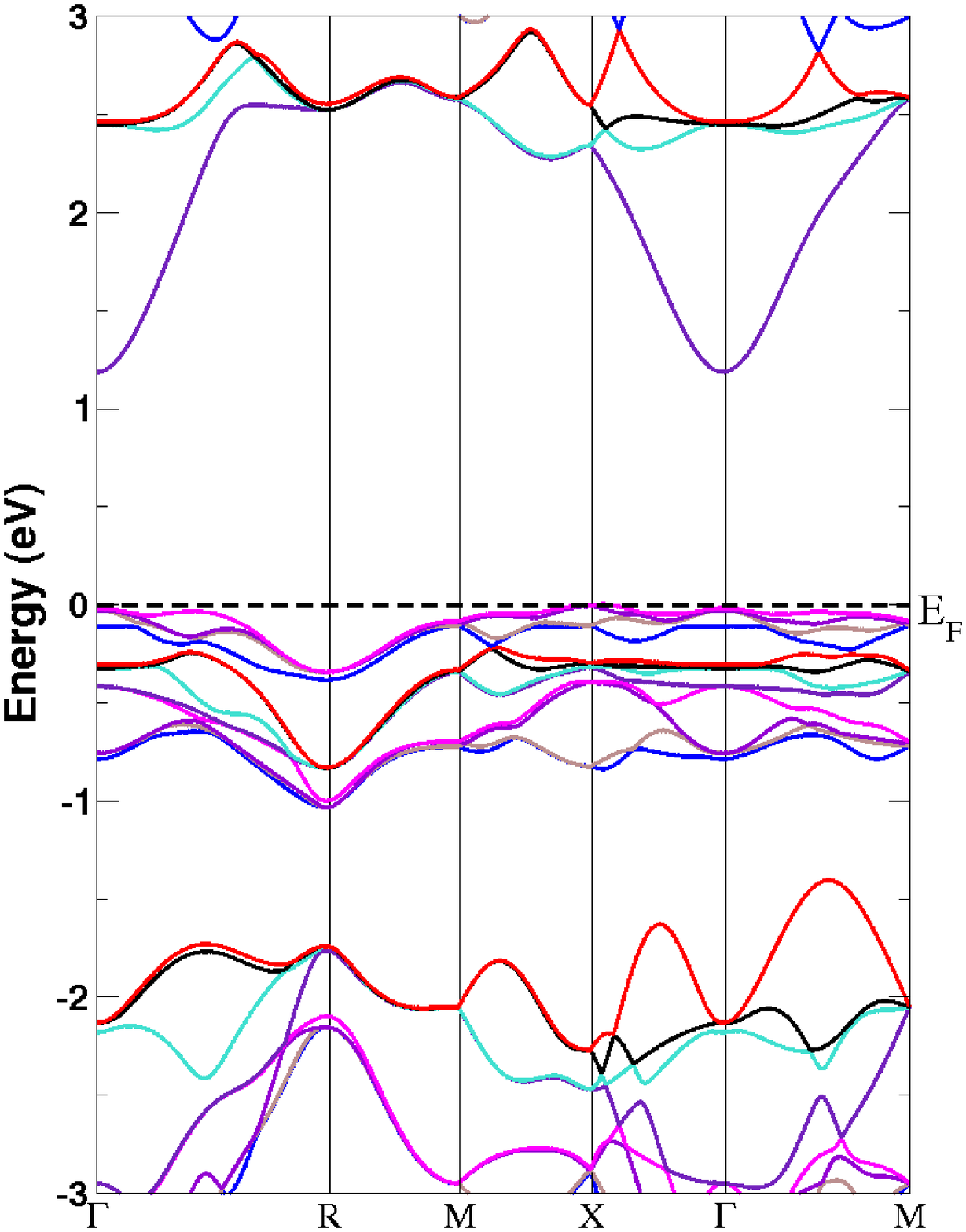}}
\caption{(Color online) Band structure of (a) Marcasite and (b) Pyrite along the high symmetry directions of the Brillouin zone}
\end{figure}

\begin{figure}
\subfigure[]{\includegraphics[width=75mm,height=75mm]{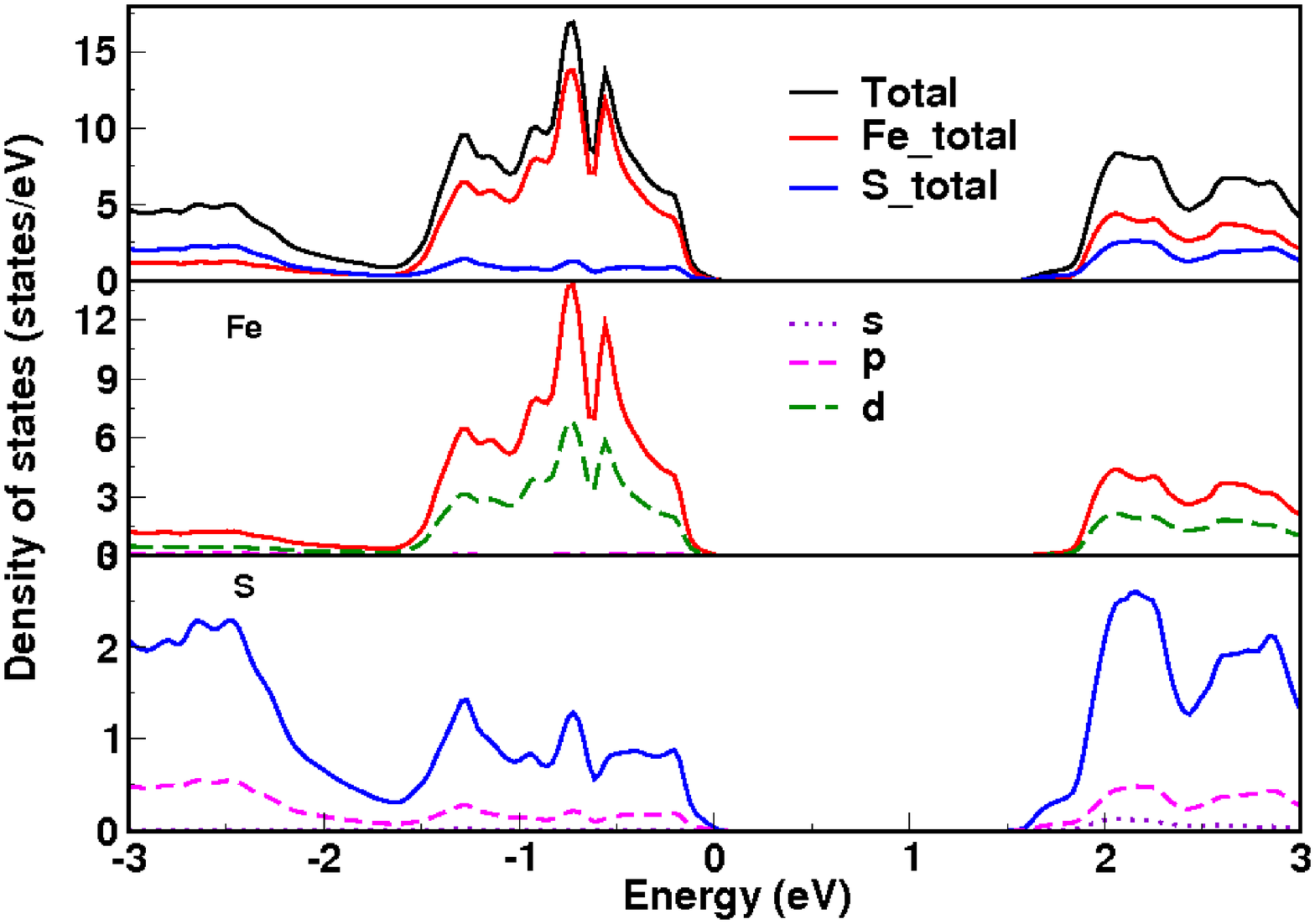}}
\subfigure[]{\includegraphics[width=75mm,height=75mm]{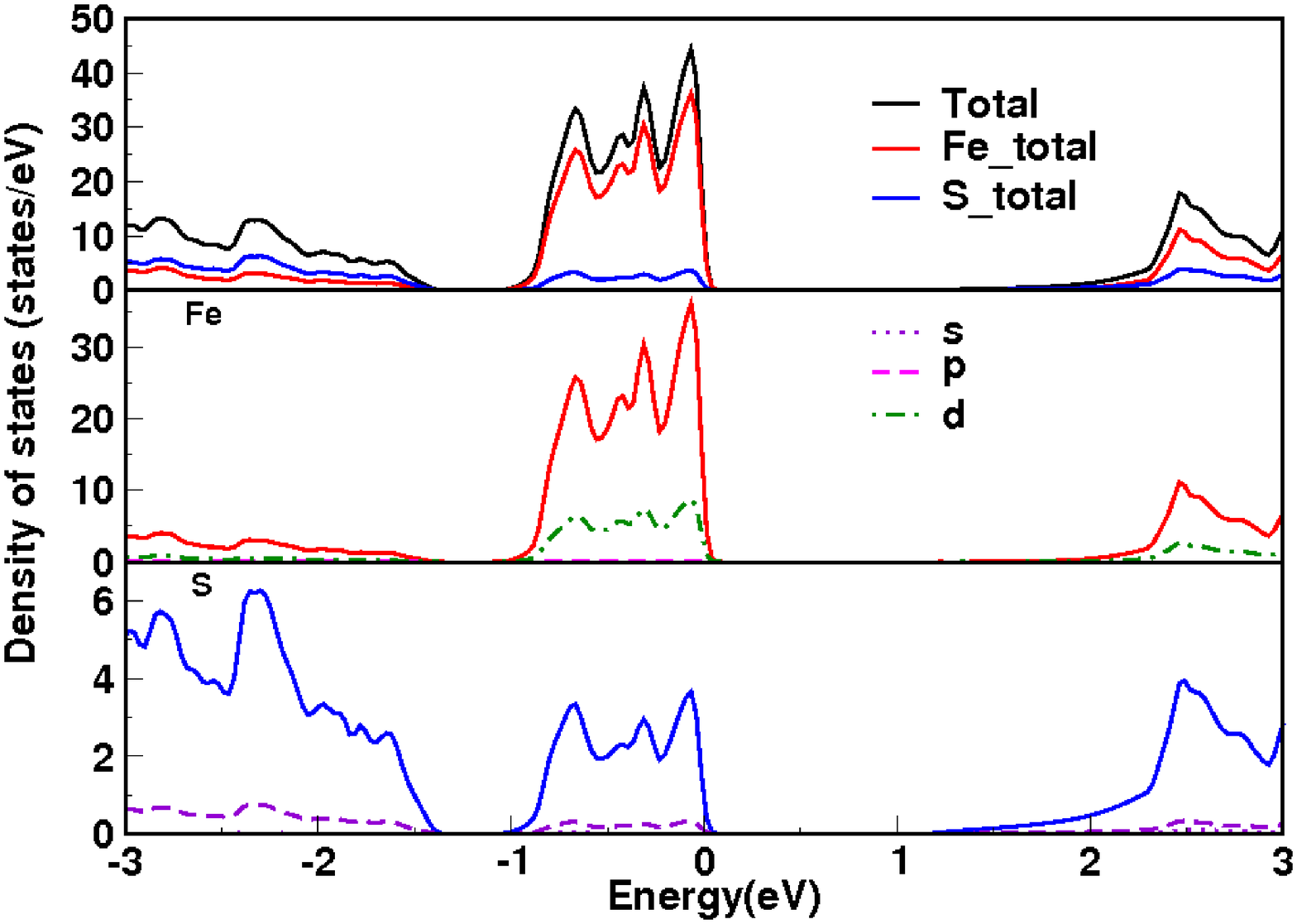}}
\caption{(Color online) Density of states of (a) Marcasite and (b) Pyrite}
\end{figure}

\begin{figure}
\subfigure[]{\includegraphics[width=75mm,height=75mm]{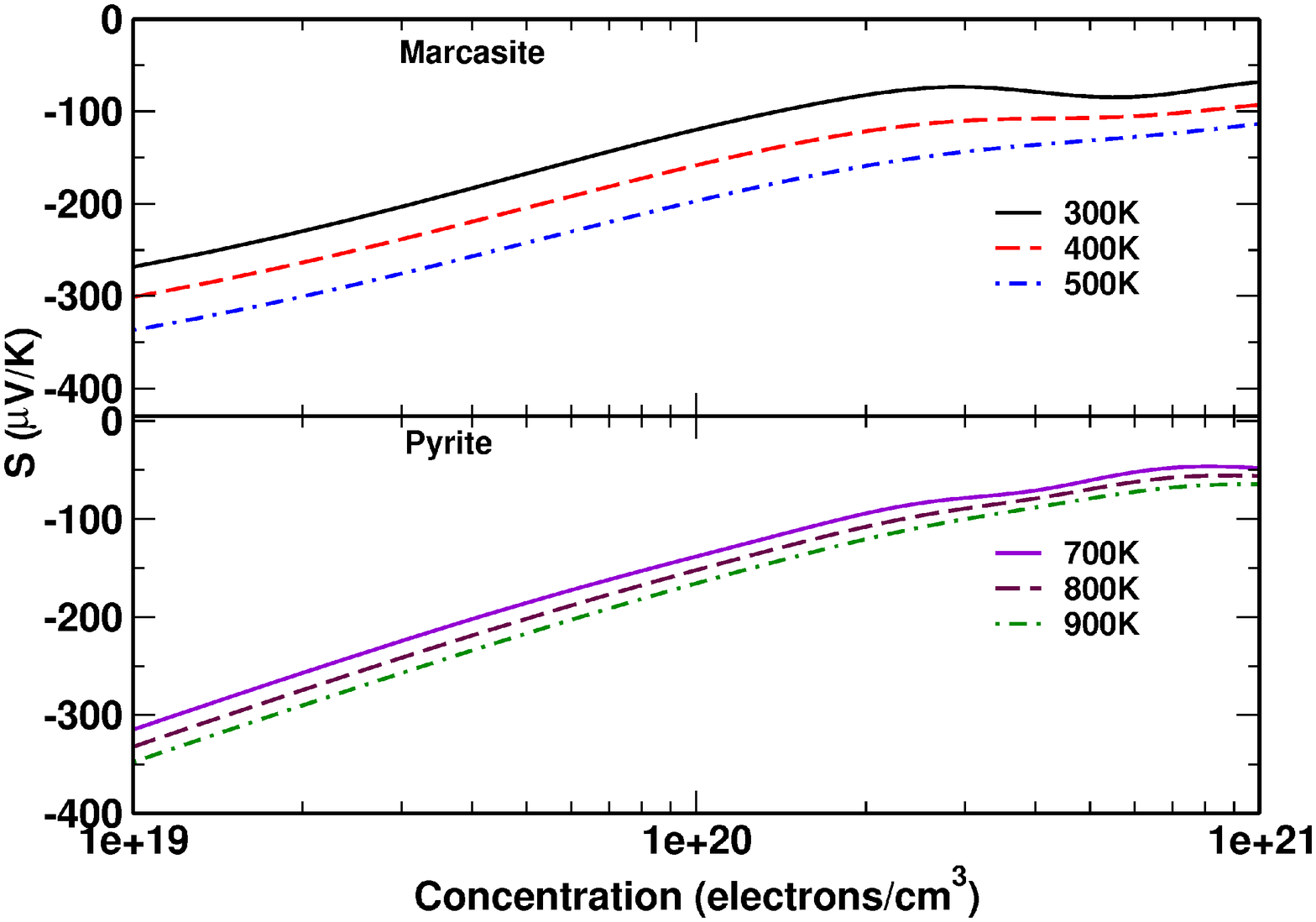}}
\subfigure[]{\includegraphics[width=75mm,height=75mm]{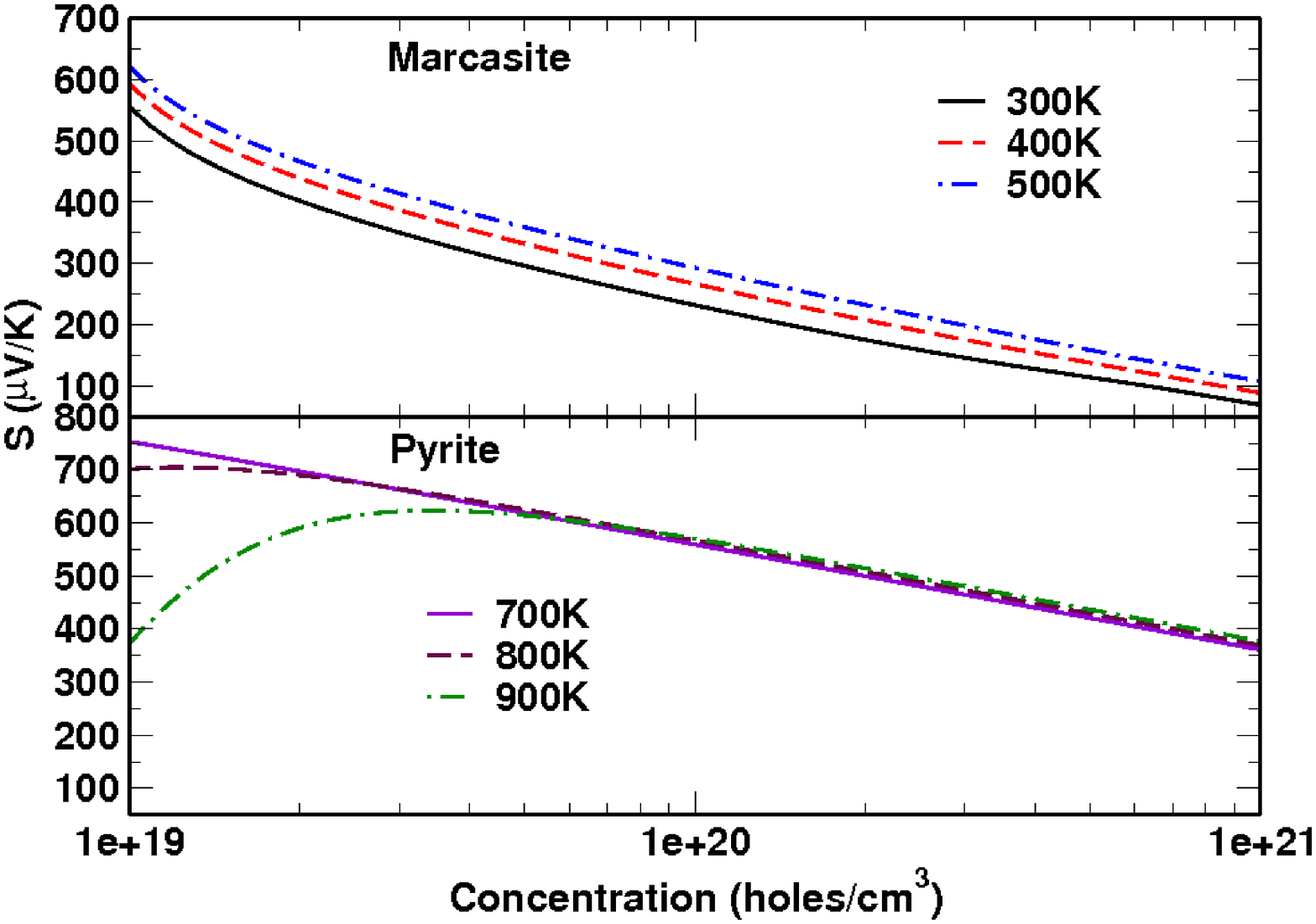}}
\caption{(Color online) Thermopower variation of marcasite and pyrite with (a) Electron concentration (b) Hole concentration}
\end{figure}

\begin{figure}
\subfigure[]{\includegraphics[width=75mm,height=75mm]{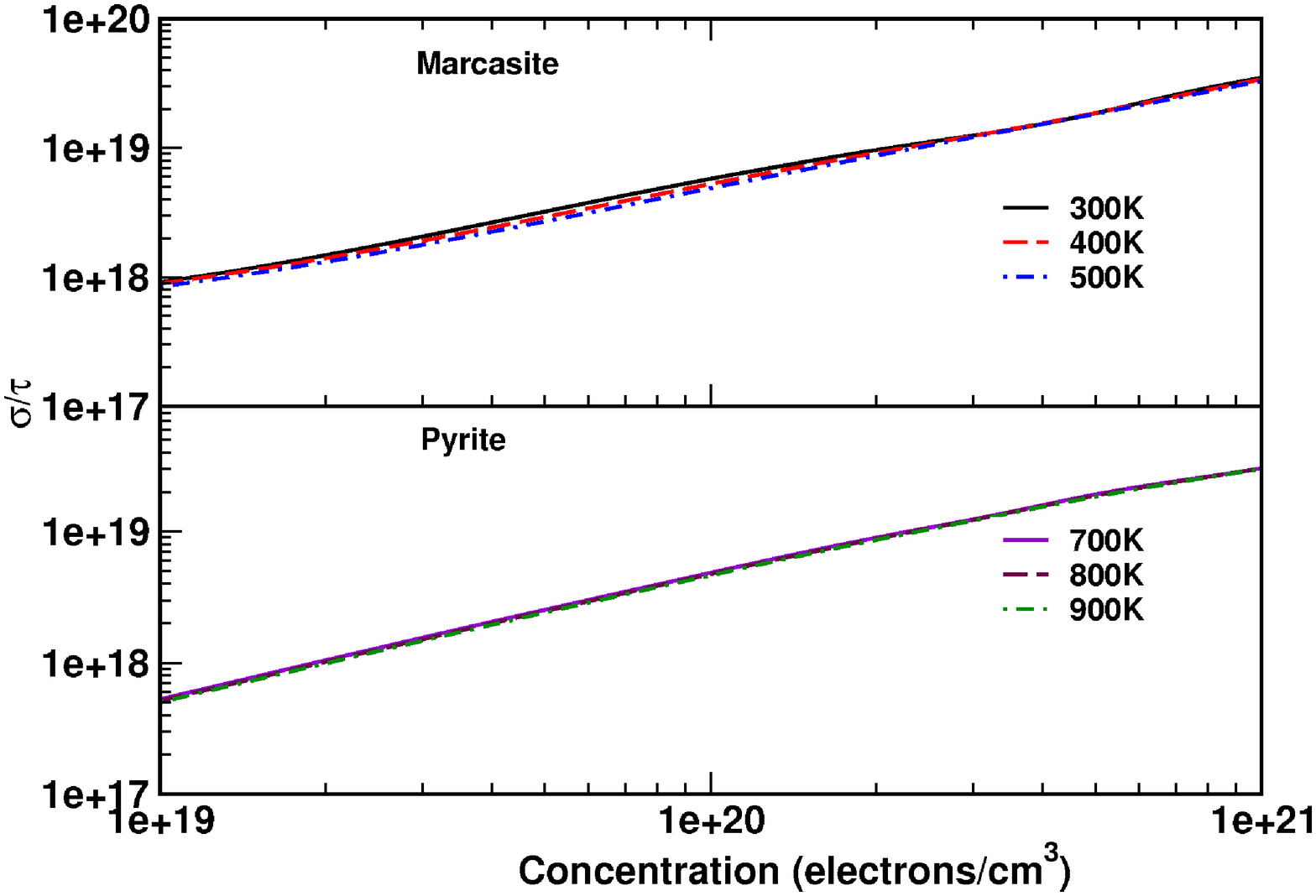}}
\subfigure[]{\includegraphics[width=75mm,height=75mm]{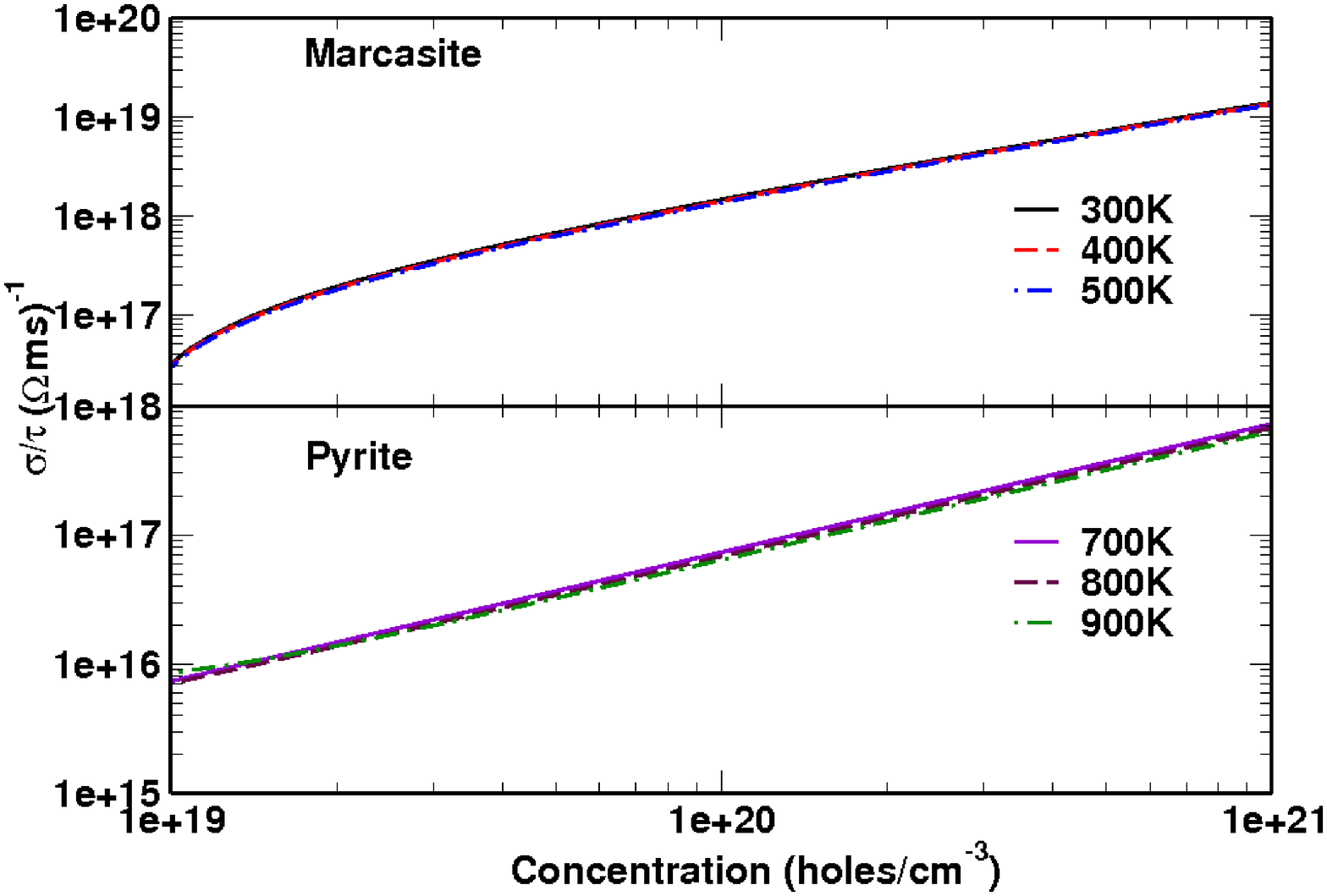}}
\caption{(Color online) Electrical conductivity variation of marcasite and pyrite with (a) Electron concentration (b) Hole concentration}
\end{figure}
\clearpage
\newpage

\begin{tocentry}
{\includegraphics[height = 3.5cm,width = 9cm]{3}}
\end{tocentry}

\end{document}